\documentclass[11pt]{article}

\usepackage{xchicago}

\newcommand {\citeAY} [1] {\citeNP {#1}}%
\newcommand {\citeAPY}[1] {\citeN  {#1}}%
\renewcommand {\showoriginalref}[1]{}
\renewcommand {\showCODEN}[1]{}
\renewcommand {\showISSN}[1]{}
\renewcommand {\showMR}[3]{}
\def\BR{{\bf R}}
\def\BA{{\bf A}}
\def\BB{{\bf B}}
\def\BL{{\bf L}}

\def\Bn{{\bf n}}
\newcommand\eq[1] {(\ref{#1})}
\newcommand{\beqa}{\begin{eqnarray}}
\newcommand{\eeqa}[1]{\label{#1}\end{eqnarray}}
\newcommand{\GO}{\Omega}
\newcommand{\Gvf}{\varphi}
\newcommand{\Real}{\mathop{\rm Re}\nolimits}

\newcommand{\Tr}{\mathop{\rm Tr}\nolimits}

\newcommand\labsect[1] {\label{sec:#1}}

\newcommand{\beq}{\begin{equation} }
\newcommand{\eeq}[1]{\label{#1}\end{equation} }
\begin{document}
\newcommand{\eps}{\epsilon}
\newcommand{\alp}{\alpha}
\newcommand{\bet}{\beta}
\newcommand{\del}{\delta}
\newcommand{\tet}{\theta}
\newcommand{\si}{\sigma}
\newcommand{\vep}{\varepsilon}
\newcommand{\vepa}[1]{\vep_{a_{#1}}}
\newcommand{\sia}[1]{\sigma_{#1}}
\newcommand{\rst}{\vec {r}}
\newcommand{\dmo}{\Delta^{-1}}
\newcommand{\bfmm}[1]{\mbox{\boldmath ${#1}$}}
\newcommand{\lmo}{\bfmm{L}_0^{-1}}
\newcommand{\jr}{\bfmm{\vec{J}}(\vec{r})}
\newcommand{\jb}{\bfmm{\vec{J}}}
\newcommand{\gmv}{\vec{\gamma}}
\newcommand{\lr}{\bfmm{ \stackrel{\leftrightarrow}{L}}(\vec{r})}
\newcommand{\lrp}{\bfmm{ \stackrel{\leftrightarrow}{L'}}(\vec{r})}
\newcommand{\lbm}{\bfmm{ \stackrel{\leftrightarrow}{L}}}
\newcommand{\gam}{{\stackrel{\leftrightarrow}{\Gamma}}{}_{1}}
\newcommand{\gamo}{{\stackrel{\leftrightarrow}{\Gamma}}{}_{0}}
\newcommand{\gamj}{{\stackrel{\leftrightarrow}{\Gamma}}{}_{1}^{j}}
\newcommand{\upsj}{{\stackrel{\leftrightarrow}{\Upsilon}}{}^{j}}
\newcommand{\kkam}[1]{\stackrel{\leftrightarrow}{K}_{#1}}
\newcommand{\kapm}[1]{\vec{\kappa}_{#1}}
\newcommand{\kaps}{\vec{\kappa}_{a_1...a_s}}
\newcommand{\kkamp}[1]{{\stackrel{\leftrightarrow}{K}}{}'_{#1}}
\newcommand{\alps}{{\stackrel{\leftrightarrow}{\alpha}_{a_1...a_s}}}
\newcommand{\blps}{{\stackrel{\leftrightarrow}{\beta}_{a_1...a_s}}}
\newcommand{\sigs}{{\stackrel{\leftrightarrow}{\sigma}}{}^{*}}
\newcommand{\als}[1]{\stackrel{\leftrightarrow}{\alpha}_{#1}}
\newcommand{\alss}[1]{\stackrel{\leftrightarrow}{\alpha}^{#1}}
\newcommand{\bls}[1]{\stackrel{\leftrightarrow}{\beta}_{#1}}
\newcommand{\As}{{\stackrel{\leftrightarrow}{A}_{a_1...a_s}}}
\newcommand{\Aa}[1]{\stackrel{\leftrightarrow}{A}_{#1}}
\newcommand{\lar}{{\stackrel{\leftrightarrow}{\Lambda}}{}_a(\vec{r})}
\newcommand{\lbr}{{\stackrel{\leftrightarrow}{\Lambda}}{}_b(\vec{r})}
\newcommand{\lalb}{{\stackrel{\leftrightarrow}{\Lambda}}{}^{l\bet}}
\newcommand{\lama}[1]{{\stackrel{\leftrightarrow}{\Lambda}}{}_{a_{#1}}}
\newcommand{\lamc}{{\stackrel{\leftrightarrow}{\Lambda}}{}_{c}}
\newcommand{\lamq}{{\stackrel{\leftrightarrow}{\Lambda}}{}_{q}}
\newcommand{\laab}{\Lambda^a_{\alp\bet}(\vec{r})}
\newcommand{\letr}{\Lambda^{l,\eta}_{\alp\bet}(\vec{r})}
\newcommand{\epr}{\bfmm{ \stackrel{\leftrightarrow} {\eps}}(\vec {r})}
\newcommand{\sr}{\bfmm{ \stackrel{\leftrightarrow} {\cal S}}(\rst,\rst~')}
\newcommand{\srst}{\bfmm{ \stackrel{\leftrightarrow} {\cal S}}(\rst,\rst~')}
\newcommand{\epb}{\bfmm{ \stackrel{\leftrightarrow} {\eps}}}
\newcommand{\epbb}{\bfmm{\eps}}
\newcommand{\cfj}[1]{{\cal F}^{#1}}
\newcommand{\cgj}[1]{{\cal G}^{#1}}
\newcommand{\Xc}[1]{{\cal X}^{#1}}
\newcommand{\Yc}[1]{{\cal Y}^{#1}}
\newcommand{\el}{\bfmm{\vec{\cal L}}}
\newcommand{\elt}{ \stackrel{\leftrightarrow}{\bfmm{\cal L}}}
\newcommand{\lb}{\bfmm{L}}
\newcommand{\lbs}{\bfmm{L}^{*}}
\newcommand{\lbsp}{\bfmm{{L}'}^{*}}
\newcommand{\lba}{\bfmm{L_a}}
\newcommand{\lbo}{\bfmm{L}_0}
\newcommand{\lbsj}{\bfmm{L}^{*j}}
\newcommand{\epa}{\bfmm{\eps_a}}
\newcommand{\epsb}{\bfmm{\eps_b}}
\newcommand{\epc}{\bfmm{\eps_c}}
\newcommand{\npa}{\bfmm{\eta_a}}
\newcommand{\npb}{\bfmm{\eta_b}}
\newcommand{\npc}{\bfmm{\eta_c}}
\newcommand{\epai}[1]{\bfmm{\eps_{a_{#1}}}}
\newcommand{\phb}{\bfmm{\phi}}
\newcommand{\psib}{\bfmm{\psi}}
\newcommand{\phbo}{\phb_0}
\newcommand{\oneb}{\bfmm{I}}
\newcommand{\onj}{I^{j}}
\newcommand{\onem}{\stackrel{\leftrightarrow}{I}}
\newcommand{\one}{\stackrel{\leftrightarrow}{\oneb}}
\newcommand{\onej}{{\stackrel{\leftrightarrow}{I}}{}^{j}}
\newcommand{\onejm}{{\stackrel{\leftrightarrow}{I}}{}^{m}}
\newcommand{\jrst}{\mbox{$\bfmm{ \vec {J}}(\rst)$}}
\newcommand{\lrst}{\el(\rst)}
\newcommand{\gle}{\gam\lmo\epb}
\newcommand{\est}{\bfmm{ \vec {E}}_0}
\newcommand{\jjst}{\bfmm{ \vec {J}}_0}
\newcommand{\ebvec}{\bfmm{ \vec {E}}}
\newcommand{\seb}{\bfmm{ \vec {e}}}
\newcommand{\sjb}{\bfmm{ \vec {j}}}
\newcommand{\diver}{\vec{\nabla}\cdot}
\newcommand{\grad}{\vec{\nabla}}
\newcommand{\rota}[1]{R_{#1}(\vec{r})}
\newcommand{\trota}[1]{R^T_{#1}(\vec{r})}
\newcommand{\xlr}{\chi_l(\vec{r})}
\newcommand{\efl}{\stackrel{\leftrightarrow}{\lbs}}
\newcommand{\eflp}{\stackrel{\leftrightarrow}{\lbsp}}
\newcommand{\elrst}{\elt(\rst)}
\newcommand{\Trst}{\Theta(\rst)}
\newcommand{\sfac}{(-1)^{s+1}}
\newcommand{\wj}{W^{j}}
\newcommand{\nj}{N^{j}}
\newcommand{\mj}{M^{j}}
\newcommand{\wjl}{W^{j-1}}
\newcommand{\yj}{Y^{j}}
\newcommand{\efls}[1]{{\stackrel{\leftrightarrow}{\bfmm{L}}}{}^{#1}}
\newcommand{\mej}[1]{{\stackrel{\leftrightarrow}{M}}{}^{j}_{#1}}
\newcommand{\mei}{{\stackrel{\leftrightarrow}{M}}{}^{i}}
\newcommand{\yej}{{\stackrel{\leftrightarrow}{Y}}{}^{j}}
\newcommand{\wej}[1]{{\stackrel{\leftrightarrow}{W}}{}^{j}_{#1}}
\newcommand{\wejl}[1]{{\stackrel{\leftrightarrow}{W}}{}^{j-1}_{#1}}
\newcommand{\norj}{{\stackrel{\leftrightarrow}{N}}{}^{j}}
\newcommand{\qej}[1]{{\stackrel{\leftrightarrow}{Q}}{}^{j}_{#1}}
\newcommand{\uej}{{\stackrel{\leftrightarrow}{U}}{}^{j}}
\newcommand{\vej}{{\stackrel{\leftrightarrow}{V}}{}^{j}}
\newcommand{\cej}{{\stackrel{\leftrightarrow}{C}}{}^{j}}
\newcommand{\dej}{{\stackrel{\leftrightarrow}{D}}{}^{j}}
\newcommand{\cxej}{{\stackrel{\leftrightarrow}{X}}{}^{j}}
\newcommand{\wea}[1]{{\stackrel{\leftrightarrow}{W}}{}^{#1}_{a}}
\newcommand{\weq}{{\stackrel{\leftrightarrow}{W}}{}^{m}_{q}}
\newcommand{\qea}[1]{{\stackrel{\leftrightarrow}{Q}}{}^{#1}_{a}}
\newcommand{\mea}[1]{{\stackrel{\leftrightarrow}{M}}{}^{#1}_{a}}
\newcommand{\ue}[1]{{\stackrel{\leftrightarrow}{U}}{}^{#1}}
\newcommand{\ve}[1]{{\stackrel{\leftrightarrow}{V}}{}^{#1}}
\newcommand{\xe}[1]{{\stackrel{\leftrightarrow}{X}}{}^{#1}}
\newcommand{\om}[1]{{\stackrel{\leftrightarrow}{I}}{}^{#1}}
\newcommand{\nm}[1]{{\stackrel{\leftrightarrow}{N}}{}^{#1}}
\newcommand{\tes}{\stackrel{\leftrightarrow}{\bfmm{T}}}
\newcommand{\brt}{\stackrel{\leftrightarrow}{B}}
\newcommand{\brst}{{\stackrel{\leftrightarrow}{B}}{}(\rst,\rst~')}
\newcommand{\aej}[1]{\vec{a}_{#1}}
\newcommand{\bej}[1]{\vec{b}_{#1}}
\newcommand{\xej}{\vec{x}}
\newcommand{\yyej}[1]{\vec{y}_{#1}}
\newcommand{\eej}{\vec{e}}
\newcommand{\pej}{\vec{p}}
\newcommand{\yest}[1]{{\stackrel{\leftrightarrow}{\bfmm{Y}}}{}^{#1}}
\newcommand{\wejz}[1]{{\stackrel{\leftrightarrow}{W}}{}^{0}_{#1}}
\newcommand{\wejo}[1]{{\stackrel{\leftrightarrow}{W}}{}^{1}_{#1}}
\newcommand{\mejo}[1]{{\stackrel{\leftrightarrow}{M}}{}^{1}_{#1}}
\newcommand{\mejt}[1]{{\stackrel{\leftrightarrow}{M}}{}^{2}_{#1}}
\newcommand{\qejo}[1]{{\stackrel{\leftrightarrow}{Q}}{}^{1}_{#1}}
\newcommand{\rprperp}{{\stackrel{\leftrightarrow}{R}}{}_{\bot}}
\newcommand{\rpj}[1]{{\stackrel{\leftrightarrow}{R}}{}^{#1}_{\bot}}

\title{The response of linear inhomogeneous systems to coupled fields: Bounds and perturbation expansions}%
\index{coupled field problems}

\author{Mordehai Milgrom$^{*}$ and Graeme W. Milton$^{**}$}
\date{$^{*}$\small{Department of Particle Physics and Astrophysics, 
Weizmann Institute of Science, 76100 Rehovot, 
Israel}\\ 
$^{**}$\small{Department of Mathematics, University of Utah, Salt Lake City, UT 84112, USA}\\
\small{emails: moti.milgrom@weizmann.ac.il, milton@math.utah.edu}}
\maketitle
\begin{abstract}
We consider the response of a multicomponent body to $n$ fields, such as electric fields, magnetic fields, 
temperature gradients, concentration gradients, etc., where each component, which
is possibly anisotropic, may cross couple the various fields with different fluxes, such as
electrical currents, electrical displacement currents, magnetic induction 
fields, energy fluxes, particle fluxes, etc. We obtain the form of the perturbation expansions of the
fields and response tensor in powers of matrices which measure the difference between each component
tensor and a homogeneous reference tensor%
\index{reference tensor}
$\BL_0$. For the case of a statistically homogeneous or periodic
composite the expansion coefficients can be expressed in terms of positive semidefinite normalization matrices alternating
with positive semidefinite weight matrices, which at each given level sum to the identity matrix. In
an appropriate basis the projection operators
onto the relevant subspaces can be expressed in block tridiagonal
form, where the blocks are functions of these weight and normalization matrices. This leads to continued fraction
expansions for the effective tensor, and by truncating the continued fraction at successive levels one obtains a
nested sequence of bounds on the effective tensor 
incorporating successively more weight and normalization matrices.  The weight matrices and normalization matrices
can be calculated from the series expansions of the fields which solve the conductivity problem alone,
without any couplings to other fields, and then they can be used to obtain the solution for the fields and effective
tensor in coupled field problems in composites.
\end{abstract}
\section{Introduction}
\setcounter{equation}{0}
This Chapter 9 of the book ''Extending the Theory of Composites to other Areas of Science'', edited by Graeme W. Milton, is concerned with the response of coupled fields and fluxes in a 
three-dimensional body $\Omega$ to potentials
prescribed at the boundary of the body and with how this response
depends on the material constants of the body. The effective tensor of a 
statistically homogeneous or periodic composite, with coupling between the fields, is a special case which we will
study in more depth. The set of fields 
$\ebvec=(\vec{E}_{1},\vec{E}_{2},...,\vec{E}_{n})$ which are each 
curl-free,%
\index{curl-free fields}
may include electric fields, magnetic fields, 
temperature gradients, or concentration gradients and the associated 
fluxes $\jb=(\vec{J}_{1},\vec{J}_{2},...,\vec{J}_{n})$ may include 
electrical currents, electrical displacement currents, magnetic induction 
fields, energy fluxes and particle fluxes. We assume there are no sources
inside the body, so each of these fluxes is divergence free.%
\index{divergence-free fluxes}
Assuming a 
simply connected topology of the body, each of the curl-free fields
derive from a potential $\vec{E}_{j}=-\grad\phi^j$. At each point
within the body or medium we 
assume a linear constitutive relation $\jb=\lbm\ebvec$ between 
the fluxes and fields through a position dependent symmetric 
positive-definite tensor $\lr$ of material constants.
The tensor may have off-diagonal couplings which cause a single driving 
field, such as a temperature gradient, to induce fluxes of all types. 

The body is assumed to be an aggregate of grains (possibly infinite in number) comprised of a 
finite number $M$ of components (phases) that have at least orthorhombic 
symmetry%
\index{orthorhombic symmetry}
with the crystal orientation varying from grain to grain,
thus $\lr $ is assumed to be piecewise constant. Let $\lb^{l,\alp}$,
$~l=1,2,\ldots M$, $\alp=1,2,3$ be the $n\times n$ principal response matrices
of the $l$-th component, defined more precisely in the next Section.

We investigate the response of the set of fluxes, $\jrst$,
measured at a given position $\rst$ within the body, to the 
potentials $\phb(\rst)=[\phi^{1}(\rst),...,\phi^{n}(\rst)]$ 
prescribed at the boundary of the body. Without loss of 
generality (see \citeAPY{Milgrom:1990:LRG} for a discussion of this point) it is 
assumed that the prescribed potentials are all in proportion to a fixed 
scalar function $f(\vec {r})$ defined at points $\vec {r}$
on the surface of the body, i.e. $\phi^{j}(\vec {r})=\phi_0^{j}f(\vec {r})$ for all $\vec {r}\in\partial\Omega$, 
and we consider how the set of 
fluxes $\jrst$ at $\rst$ vary with the choice of the vector 
 $\phbo=(\phi^{1}_{0},...,\phi^{n}_{0})$ of proportionality constants: since this relation is linear it is governed by a
response tensor%
\index{response tensor}
$\lrst$ giving $\jrst=\lrst\phbo$. This tensor $\lrst$
is the object of our analysis. Specifically
we examine the dependence of $\lrst$ on the set of crystal moduli%
\index{crystal moduli}
$\lb^{l,\alp}~~(l=1,2,...,M,\alp=1,2,3)$ 
when each is close to a constant tensor $\lbo$,
i.e. when the material constants of the body are close to being 
homogeneous and isotropic. To simplify notations these crystal moduli
are relabeled as $\lb_{a}~~(a=1,2,...,p)$, avoiding repetitions in 
the original set of crystal moduli due to crystal symmetries of isotropy
or uniaxiality: thus, when there are no symmetries (other than orthorhombic 
symmetry) $a$ represents the pair $(l,\alp)$ and $p=3M$, but $p$ could be less than $3M$
if some of the phases are isotropic or uniaxial.

A formal expression is obtained for the coefficients appearing in the 
series expansion%
\index{series expansions}
of $\lrst$ in powers of the differences 
$\epa=\lb_{a}-\lbo~(a=1,2,..,p)$. We say formal 
because these coefficients are difficult to evaluate and because their
(nonlinear and nonlocal) dependence on
the overall shape of the body, on the division of the body into
grains and on the orientation of the crystals in each grain is
complicated. What is interesting is the explicit form of the
expansion. This is a non trivial issue since the set of matrices
$\lb_{a}~(a=1,2,...,p)$
do not necessarily commute. The issue has been addressed in part by
\citeAPY{Milgrom:1990:LRG} 
from general analytic considerations. Milgrom noted that the
functional dependence of $\lrst$ on the $\lb_{a}$ must satisfy two
constraints:

(i){\it Covariance},%
\index{covariance}
the property that for any real, nonsingular, $n$ by $n$ matrix
$\bfmm{W}$ with transpose $\bfmm{W}^{T}$ acting only on the field 
indices, the response tensor $\lrst$ transforms to 
$\bfmm{W}\lrst\bfmm{W}^{T}$ when all of the
crystal moduli $\lb_{a}$ are replaced by the moduli 
$\bfmm{W}\lb_{a}\bfmm{W}^{T}$. Covariance follows from the observation
that we are free to define a new set of (curl-free) fields 
$\ebvec~'=(\bfmm{W}^{T})^{-1}\ebvec$ and a new set of (divergence-free) fluxes 
$\jb~'=\bfmm{W}\jb$ by taking linear combinations of the
old set of fields and fluxes while preserving at the same time
the self-adjointness of the tensor $\lr'=\bfmm{W}\lr\bfmm{W}^{T}$ 
in the constitutive relation $\jb~'=\lbm'\ebvec~'$. 
Clearly $\bfmm{W}\lrst\bfmm{W'}$ is simply
the old response tensor $\lrst$ expressed in terms of the new fields.

(ii){\it Disjunction},%
\index{disjunction}
the property that when the matrices $\lb_{a}$ are
block diagonal of the same form then so must $\lrst$ have a similar
block diagonal form in the
field indices, and furthermore the elements of $\lrst$ within
each block only depend on the elements of the $\lb_{a}$'s in
the corresponding blocks. Disjunction follows from the observation
that if a subset of fields is decoupled from another subset of
fields then the effective response tensor must reflect this
decoupling.
\par
These analytic considerations alone eliminate from consideration
many candidates for the terms in the series expansion, such as for
example $\lbo^{-1}\epai{1}\epai{2}$, and leave terms such as
$\epai{1}\lmo\epai{2}\lmo\epai{3}$ (that in fact do occur in the
series expansion) as natural candidates. 


The technique we employ in the present Chapter is a simple generalization of an
approach used in the theory of composite materials to
derive series expansions for the effective conductivity or
elasticity tensor  of a nearly homogeneous multiphase material. A lot
of the progress that has been made on series expansions
and associated bounds%
\index{bounds!effective tensors}
on effective tensors is summarized in the books of  \citeAPY{Cherkaev:2000:VMS},
\citeAPY{Milton:2002:TOC}, \citeAPY{Allaire:2002:SOH}, \citeAPY{Torquato:2002:RHM}, \citeAPY{Tartar:2009:GTH}.
\citeAPY{Brown:1955:SMP}, in a pioneering paper, obtained the series
expansion of the effective conductivity $\sigma^{*}$ of an
isotropic composite of two isotropic components with nearly
equal conductivities $\sigma_{1}$ and $\sigma_{2}$, and found
that the coefficient of $(\sigma_{1}-\sigma_{2})^{n}$ in
this expansion depends on
the $n$-point correlation function%
\index{correlation function!n@$n$-point}
giving the probability that a fixed
configuration of $n$-points lands with all points in component 1 when placed
randomly in the composite. Subsequently many other
series expansions%
\index{series expansions}
were derived for the effective conductivity
tensor or elasticity tensor of nearly homogeneous
composites:%
\index{composites!nearly homogeneous}
see for example, \citeAPY{Herring:1960:ERI}, \citeAPY{Prager:1960:DIM},
\citeAPY{Beran:1963:SPE}, \citeAPY{Beran:1968:SCT},\citeAPY{Beran:1970:MFV}), \citeAPY{Fokin:1969:CEE}, \citeAPY{Dederichs:1973:VTE}, \citeAPY{Hori:1973:STE}, \citeAPY{Zeller:1973:ECP}, \citeAPY{Gubernatis:1975:MEP}, \citeAPY{Kroner:1977:BEE}, \citeAPY{Willis:1981:VRM},
\citeAPY{Milton:1982:NBE}, \citeAPY{Phan-Thien:1982:NBE}, \citeAPY{Sen:1989:ECA}, \citeAPY{Torquato:1997:EST},
Tartar (\citeyearNP{Tartar:1989:MSA}, \citeyearNP{Tartar:1990:MNA}), and \citeAPY{Bruno:1991:TEB}. Our analysis closely
follows that of \citeAPY{Willis:1981:VRM} and Phan-Thien and Milton 
(\citeyearNP{Phan-Thien:1982:NBE}, \citeyearNP{Phan-Thien:1983:NTO}).

Our analysis gives, as a simple corollary, a series expansion
for the effective tensor $\efl$ that governs the constitutive
relation between the local average of $\jb$ and
the local average of $\ebvec$ in a statistically homogeneous or
periodic composite material.%
\index{composites!statistically homogeneous}%
\index{composites!periodic}
(These averages are taken over a length
scale much larger than the microstructure, yet smaller than any 
macroscopic lengths associated with variations in the applied fields.) 
This expansion is derived in Section 4 where the body is assumed to
be filled with such a composite material, with microstructure much
smaller than the dimensions of the body. From the response tensor $\lrst$ 
associated with linear potentials specified on the boundary, i.e. with
$f(\vec {r})=-\vec {r} \cdot \vec {v}_{0}$ on $\partial\Omega$, where $\vec {v}_{0}$ 
gives the direction of the applied field, we directly obtain the effective
tensor $\efl$ of the composite. 

The coefficients in the series expansion of $\efl$ in powers of 
the $\epa~(a=1,2,..,p)$ are useful for obtaining bounds on $\efl$.%
\index{bounds!coupled field problems}
In particular they likely contain sufficient information 
to determine the weight%
\index{weight matrices}
and normalization matrices%
\index{normalization matrices}
that were introduced
by Milton (\citeyearNP{Milton:1987:MCEa}, \citeyearNP{Milton:1987:MCEb}),
following the introduction of scalar valued weights and normalization factors
by \citeAPY{Milton:1985:TCC}. Thus these parameters are seen to have a natural
significance in the context of coupled field problems.%
\index{coupled field problems}
In any case the weight matrices and normalization matrices
can be calculated from the series expansions of the fields. It is noteworthy
that they can be calculated from the series expansions of the fields which solve the conductivity problem alone,
without any couplings to other fields, and then they can be used to obtain the solution for the fields and effective
tensor  $\efl$ in coupled field problems.

With these geometric parameters 
we show how one can compute, for coupled field problems, the Wiener-Beran%
\index{bounds!Wiener-Beran type} 
and Hashin-Shtrikman type bounds%
\index{bounds!Hashin-Shtrikman type}
of any order: these
bounds, derived for the effective conductivity by 
\citeAPY{Milton:1981:BTO} and \citeAPY{Milton:1982:CTM}
(see also \citeAY{McPhedran:1981:BET}) and extended 
here to bounds on $\efl$, generalize the bounds of \citeAPY{Wiener:1912:TMF},
\citeAPY{Hashin:1962:VAT}, \citeAPY{Beran:1965:UVA}, \citeAPY{Willis:1977:BSC}, 
\citeAPY{Phan-Thien:1982:NBE}, and \citeAPY{Sen:1989:ECA}. They do not, however,
encompass the optimal two-dimensional, two-phase bounds of 
\citeAPY{Cherkaev:1992:ECB} and \citeAPY{Clark:1995:OBC} which couple 
effective tensors
using additional information about the 
differential constraints on the fields, or duality relations%
\index{duality relations}
satisfied by
the effective tensor as a function of the component moduli.%
\index{effective tensor!as a function of the component moduli}
Again many of the existing bounds are summarized in the books of 
 \citeAPY{Cherkaev:2000:VMS}, \citeAPY{Milton:2002:TOC}, \citeAPY{Allaire:2002:SOH}, \citeAPY{Torquato:2002:RHM}, \citeAPY{Tartar:2009:GTH}.
\section{Setting of the problem and equations for the fields}
\setcounter{equation}{0}
We consider the problem of linear response to $n$ coupled fields
derivable from potentials $\phi^k,~k=1,...,n$. The problem is
described, in detail, by \citeAPY{Milgrom:1990:LRG},
 and we give a succinct
description here.
 The body consists
of a space domain $\Omega$ within which the position-dependent
response tensor is $L_{\alp i\bet k}(\vec {r})$, where $i,k$ are field indices%
\index{field indices}
and $\alp,\bet$ are space indices.%
\index{space indices}
The $\alp$th component of the $i$th
flux is given by the constitutive relation
\beq J^i_\alp(\vec {r})=-\sum_{k=1}^n\sum_{\beta=1}^3L_{\alp i\bet k}(\vec {r})\partial_\bet\phi^k
(\vec {r}), \eeq{Kflux}
or, suppressing the indices:
\beq \jb=-\lbm\grad\phb. \eeq{Kfluxco}
We shall be using boldface letters for quantities that are vectors
or tensors in the field indices. Also, a $\rightarrow$ above a character
indicates a vector in the space indices and a $\leftrightarrow$ above a 
character indicates a matrix in the space indices.
 So, for example, $\jb,~\ebvec,$ and $\seb$
are vectors in both space and field indices; $\lbm$, and $\epb$ are second
rank tensors in both types of indices; $\lb$ is a matrix 
in the field indices; $\phb$ and $\phbo$ are vectors in the field 
indices; $\vec{r}$ is a vector in the space
indices; and $\gam$ is a matrix in the space indices. 

The equation
\beq \diver {\bfmm{ \vec {J}}}=0, \eeq{Kconst}
determines  the fields $\phb$ within $\Omega$, given the boundary conditions.

The response we consider is the field vector of
$n$ fluxes, $\jrst,$  measured at a given position $\rst$ within $\Omega$,
 and is taken to respond to the boundary conditions
dictated on the surface, $\partial\Omega$, of $\Omega$.
 As explained in \citeAPY{Milgrom:1990:LRG}, we may, without loss of
generality, restrict ourselves to boundary conditions of the form
\beq \phb(\vec {r})=\phbo f(\vec {r}),~~~~
\vec {r}\in \partial\Omega. \eeq{Kbound}
We then define the response matrix%
\index{response matrix}
$\lrst$ such that
\beq \jrst=\lrst\phbo. \eeq{Kresp}
We shall be interested in a piecewise-homogeneous system,
so $\Omega$ is divided into a (possibly infinite) number of domains, as in Fig. 1,
 each of which is
filled with one of $M$ (possibly anisotropic) components, with an arbitrary
orientation of its axes. We restrict ourselves to components that have,
at least, an orthorhombic symmetry.%
\index{orthorhombic symmetry}
The response matrix, $\lrst,$ depends, then,
 on the shape of $\Omega$, on the choice of $f(\vec {r})$, on the 
division of $\Omega$ into sub-domains, on the orientations
of the different components within these homogeneous sub-domains, and on the
response properties of the individual components.
In the principal axes of the $l$th component we can write
\beq L^l_{\alp k\bet m}=L^{l,\alp}_{km}\del_{\alp\bet}, \eeq{Kprinc}
where there is no summation over $\alpha$ and the 
$\lb^{l,\alp}~~(\alp=1,2,3)$ are
 the principal response matrices%
\index{principal response matrices}
of component
$l$, $l=1,2,\ldots,M$. Let $p$ be the total number of such principal matrices characterizing
all the components. So, there is only one such matrix 
for an isotropic component,
two for a component with uniaxial symmetry,%
\index{uniaxial symmetry}
and three for a component
with orthorhombic symmetry.%
\index{orthorhombic symmetry}
We shall use a single index notation with
$\lba,~~a=1,...,p$ instead of the doubly indexed $\lb^{l,\alp}.$
(Depending on the symmetry, $\alp$ here takes one, two, or three values.)
\par
In the isotropic and homogeneous case we have
\beq \lba=\lbo, \eeq{Khom}
for all $a$. When there are departures from isotropy and homogeneity
we write
\beq \lba=\lbo+\epa ,  \eeq{Kllee}
and seek to expand $\lrst$ in the elements $\eps^a_{ik} $ of the $\epa$'s.
\par
To this end we first derive a formal expression for the driving field,
 $\ebvec=-\grad \phb$ produced within $\Omega$ by the boundary conditions $\phbo$.
Let $\est$ be the driving field that is produced by these same boundary
conditions {\it in the homogeneous, isotropic} case.
We can write
\beq \est(\vec{r})=\phbo\vec{v}_0(\vec{r}),  \eeq{Kesta}
where $\vec{v}_0=-\grad\psi_{0},$ and $\psi_{0}$ is 
the single-field solution
of the Laplace equation, in $\Omega$, with boundary condition
 $\psi(\vec {r})=f(\vec {r})$ on $\partial\Omega$; thus 
$\diver\est=0.$ The difference field
\beq \seb\equiv\ebvec-\est=-\grad\bfmm{\psib},  \eeq{Ksmalle}
is derivable from a potential $\psib$, that vanishes on $\partial\Omega$. Now introduce
\beq \epb(\vec{r})=\lr-\lbo\onem,  \eeq{Kdifffield}
where we use $\onem$ for the unit matrix in space indices;
$\one$ for the identity in both space and field indices; and $I$ for the 
identity operator which when acting on a function leaves it invariant. Then the flux field,
\beq \jr=\lr\ebvec(\vec{r}), \eeq{Krela}
can thus be written as
\beq \jr=[\lbo\onem+\epb(\vec{r})](\est+\seb)(\vec{r}). \eeq{Kjr}
Taking the divergence of (\ref{Kjr}), and remembering that $\jb$ and
$\lbo\est$ are divergence-free, we obtain
\beq \lbo\diver\seb+\diver(\epb\ebvec)=0,  \eeq{Kgga}
or equivalently,
\beq \Delta\bfmm{\psib}=\diver(\lmo\epb\ebvec).  \eeq{Kggaa}
Define, now, the inverse-Laplacian,%
\index{inverse-Laplacian}
$\dmo,$ as the  nonlocal operator
which, acting on a density function
$\rho(\vec{r})$, defined in $\Omega$,
gives the potential $\varphi$ that solves the Poisson's equation%
\index{Poisson's equation}
$\Delta\varphi=-\rho$, and vanishes on the surface, $\partial\Omega$.
Then, from (\ref{Kggaa}) and (\ref{Ksmalle}) we can write
\beq \seb=-\gle\ebvec, \eeq{Ksebb}
where
\beq \gam\equiv\grad\dmo\diver, \eeq{Kgama}
is nonlocal, with kernel $\gam(\vec{r},\vec{r}~')$, and acts on a vector field $\vec{u}(\vec{r}~')$ to give 
the vector field 
\beq \vec{v}(\rst)= \int_{\Omega} d\rst~'~ \gam(\rst~,\rst')\vec{u}(\rst~'), 
\eeq{Kdgam} 
that has
the same divergence as $\vec{u}$, and is derivable from a potential that
 vanishes on $\partial\Omega$. Clearly, $\gam$ is a
projection operator:%
\index{projection operators!gz@$\gam$}
\beq  \gam\gam=\gam, \eeq{Kproj}
implying its kernel satisfies
\beq \gam(\vec{r},\vec{r}~')=\int_{\Omega} d\rst~''~~\gam(\vec{r},\vec{r}~'')\gam(\vec{r''},\vec{r}~'). \eeq{Kernproj}
In addition, because $\gam$ gives zero when it acts
on a uniform vector field and always produces a vector field with zero
integral over $\Omega$, we have
\beq  \int_{\Omega} d\rst~'~ \gam(\vec{r},\vec{r}~')=0,~~~~~~
 \int_{\Omega} d\rst~ \gam(\vec{r},\vec{r}~')=0. \eeq{Kinte}
The operator $\gam$ is also self-adjoint, i.e. 
\beq \gam(\vec{r},\vec{r}~')=[\gam(\vec{r}~',\vec{r})]^T, \eeq{Kgsad0}
where $T$ denotes the transpose. To see this, suppose one is given vector fields $\vec{u}(\rst)$ and $\vec{v}(\rst)$. Let $\Gvf(\rst)$ 
and $\psi(\rst)$ be potentials that vanish on the boundary $\partial\Omega$ such that 
\beq \vec{u}=\grad\Gvf+\grad\times\BA,\quad \vec{v}=\grad\psi+\grad\times\BB, \eeq{Kgsad1}
for some vector potentials $\BA(\rst)$ and $\BB(\rst)$. Then the definition of $\gam$ implies $\gam\vec{u}=\grad\Gvf$ and $\gam\vec{v}=\grad\psi$. So we have
\beq \int_{\GO}\vec{v}\cdot(\gam\vec{u})=\int_{\GO}\grad\psi\cdot\grad\Gvf+\int_{\GO}(\grad\times\BB)\cdot\grad\Gvf. \eeq{Kgsad2}
Using the divergence theorem, the last integral vanishes,
\beq \int_{\GO}(\grad\times\BB)\cdot\grad\Gvf=\int_{\GO}\diver[\Gvf(\grad\times\BB)]=\int_{\partial\GO}\Gvf\Bn\cdot(\grad\times\BB)=0,
\eeq{Kgsad3}
where $\Bn$ is the outwards normal to $\partial\GO$, and we have used the fact that $\Gvf=0$ on $\partial\GO$. Switching the roles of $\vec{v}$ and $\vec{u}$ in \eq{Kgsad2} gives the same result, and so we obtain
\beq \int_{\GO}\vec{v}\cdot(\gam\vec{u})=\int_{\GO}\vec{u}\cdot(\gam\vec{v}),
\eeq{Kgsad4}
which means $\gam$ is self-adjoint.

Adding $\est$ to both sides of (\ref{Ksebb}), we can write
\beq (I+\gle)\ebvec=\est,  \eeq{Kferel}
or
\beq \ebvec=(I+\gle)^{-1}\est.  \eeq{Kfield}
Thus, from the definition of the response matrix $\lrst$, equation (\ref{Kresp}), from
relation (\ref{Kesta}) between $\est$ and $\phbo$, and from 
relation (\ref{Krela}) between $\jb$ and $\ebvec$, we get
\beq \lrst=\int d\rst~'~\srst\vec{v}_0(\rst~'),  \eeq{Kconv}
where $\srst$ is the kernel of a nonlocal operator ${\bfmm{ \stackrel{\leftrightarrow} {\cal S}}}$  [acting on the field $\vec{v}_0$],
given by
\beq {\bfmm{ \stackrel{\leftrightarrow} {\cal S}}}=(\lbo\onem+\epb)(I+\gle)^{-1}.  \eeq{Kexpr}
The vector field $\vec{v}_0$ only carries the information on the exact
form of the boundary conditions $[f(\rst_{0})]$; it is $\srst$ that 
plays the role of the response tensor of the system.

\section{The expansion of the response tensor}
\setcounter{equation}{0}
We now use (\ref{Kexpr}) to develop a series 
expansion%
\index{series expansions}
for the response tensor%
\index{response tensor}
$\srst$.
Specializing to the piecewise homogeneous case, we express $\epb(\vec{r})$
in terms of the $\epa$'s defined in equation (\ref{Kllee}).
Defining the indicator function,%
\index{indicator function}
$\xlr$, such that $\xlr=1$
 in a subregion occupied by component $l$, $l=1,2,\ldots,M$ and $\xlr=0$
otherwise, we can write for the $\alp\bet$ element of $\epb$
\beq \epbb_{\alp\bet}(\vec{r})=\sum_{l=1}^M\sum_{\eta=1}^3\xlr\rota{\alp\eta}
\epbb^{l,\eta}\trota{\eta\bet}, \eeq{Ksusu}
where
\beq \epbb^{l,\eta}=\lb^{l,\eta}-\lbo,  \eeq{Kaa}
and $\lb^{l,\eta}$ are the principal response matrices%
\index{principal response matrices}
of component $l$,
$\rota{}$ is the rotation matrix from the principal axes to the orientation
the component has at position $\vec{r}$, and $\trota{}$ is its transpose
(inverse).
\par
Equation (\ref{Ksusu}) can be cast in the form
\beq \epbb_{\alp\bet}(\vec{r})=\sum_{a=1}^p\laab\epa, \eeq{Kbbb}
where the elements, $\laab$,
 of $\lama{}$ are defined as follows: For an orthorhombic component, there
are three $\lama{}$'s, where $a$ replaces the double index $l,\eta$,
and
\beq \letr=\xlr\rota{\alp\eta}\trota{\eta\beta} \eeq{Kccc}
(with no summation over $\eta$).
When the component $l$ is isotropic, it contributes only one $\lama{}$,
with
\beq \laab=\xlr\sum_{\eta=1}^3\rota{\alp\eta}\trota{\eta\beta}=
\xlr\del_{\alp\bet}.  \eeq{Kddd}
Similarly, for a uniaxial component there are two matrices
$\lar$. It is easy to ascertain that
\beq \lar\lbr=\del_{ab}\lar,  \eeq{Keee}
and we also have
\beq \sum_{a=1}^{p}\lar=\onem.  \eeq{Klll}
Now, substituting (\ref{Kbbb}) in expression (\ref{Kexpr}) 
for ${\bfmm{ \stackrel{\leftrightarrow} {\cal S}}}$, and expanding, we get
\beq  {\bfmm{ \stackrel{\leftrightarrow} {\cal S}}}=\lbo\onem+\sum_{s=1}^\infty\sum_{a_1,...,a_s=1}^p\sfac\kkam{a_1...a_s}
\epai{1}\lmo\epai{2}\lmo...\lmo\epai{s},  \eeq{Kfff}
where the reduced operator%
\index{reduced operator}
$\kkam{a_1...a_s}$ is given by
\beq \kkam{a_1...a_s}=(I-\gam)\lama{1}\gam\lama{2}\gam...\gam\lama{s}.
 \eeq{Kggg}
{\it Note that each operator $\gam$ in the above relation acts on the whole
expression to its right including the field on which $\kkam{a_1...a_s}$ acts: it does not just act on the adjacent $\lar$ factor.}
The reduced operators, which are matrices in the space indices,
are purely geometrical. They depend
on the geometry of the region $\Omega$, on its division into
 homogeneous sub-regions, and on the orientation of the components within these
sub-regions. They do not depend
 on the form of the boundary condition $f(\rst_{0})$, which enter through
$\vec{v}_0(\vec{r})$ (on which the $\kkam{}$'s act);
 they also do
 not depend on the response coefficients of the components, which enter
 through the field-matrix terms in (\ref{Kfff}).
\par
Using (\ref{Kconv}), the corresponding reduced,
expansion coefficients of the response $\lrst$
are
\beq \kaps(\rst)=[\kkam{a_1...a_s}\vec{v}_{0}](\rst)=\int_{\Omega} d\rst~'~\kkam{a_1...a_s}(\rst,\rst~')
 \vec{v}_{0}(\rst~'),\eeq{Kaaaa}
where $\kkam{a_1...a_s}(\rst,\rst~')$ is the kernel of the operator $\kkam{a_1...a_s}$.
Note that the reduced coefficients are not independent: Summing over the last
index gives
\beq \sum_{a_s=1}^{p}\kaps=0, \eeq{Kaabb}
from (\ref{Klll}), and the fact that $\gam$ acting on a
divergence-free vector field (such as $\vec{v_0}$)
gives 0. Summing over the first index we also have
\beq \sum_{a_1=1}^{p}\kaps=0, \eeq{Kaabbext}
because $(I-\gam)\gam=0$. Summing over any, but the last, or first, index gives
a reduced coefficient with one less index:
\beq \sum_{a_i=1}^{p}\kaps=
 \kapm{a_1...a_{i-1}a_{i+1}...a_s}. \eeq{Kaacc}
These follow directly from (\ref{Kggg}), and stem from the fact
that we could arbitrarily redefine $\lbo$ by adding to it a constant
matrix, and subtract that matrix from the $\epa$'s, without affecting
$\sr$. So there are really only $(p-1)^s$ independent $s$-th order coefficients, not $p^s$.
\section{The expansion of the effective tensor of a 
composite}
\setcounter{equation}{0}
We now focus attention on an important subclass of inhomogeneous
bodies: those filled with a statistically homogeneous or periodic
composite material with microstructure much smaller than the dimensions
of the body. It is well-known and can be rigorously proved (see for
example \citeAPY{Golden:1983:BEP}) that if there exists an
intermediate length scale $\lambda$ much larger than the homogeneities yet
much smaller than the length scales associated with the
dimensions of $\Omega$ and with variations in the 
applied potentials, then $\lr$ can be replaced by a constant effective
tensor $\lb^{*}$ without disturbing the macroscopic response of the
body. At distances from the boundary $\partial\Omega$, inside the
body, sufficiently greater than $\lambda$ 
this effective tensor $\efl$ governs the relation between
the fields 
\beq <\jb>_{\Trst}={\frac{1}{|\Trst|}}\int_{\Trst} d\rst~'~ \jb(\rst~'),~~~~
<\ebvec>_{\Trst}={\frac{1}{|\Trst|}}\int_{\Trst} d\rst~'~ \ebvec(\rst~'), \eeq{Kave}
obtained by averaging $\jb(\rst~')$ and $\ebvec(\rst~')$ over a sphere
$\Trst$ of volume $|\Trst|$, centered at $\rst$, with
radius $\lambda$, through the constitutive relation 
\beq <\jb>_{\Trst}= \efl<\ebvec>_{\Trst}. \eeq{Keffe}
Another tensor of interest is the microscopic response tensor%
\index{response tensor!microscopic}
$\elt(\rst~',\rst)$ which governs the linear relation between $\jb(\rst~')$, for points
$\rst~'$ in $\Trst$, and
$<\ebvec>_{\Trst}$:
\beq  \jb(\rst)= \elt(\rst~',\rst)<\ebvec>_{\Trst}. \eeq{Kdelt}
This tensor $\elt(\rst~',\rst)$ is only well-defined if there is a sufficient separation of 
length scales so that homogenization theory%
\index{homogenization theory}
(see the many references in the introduction in Chapter 1 of this book, ''Extending the Theory of Composites to Other Areas of Science'' edited by Graeme W. Milton,
and in particular \citeAY{Bensoussan:1978:AAP} 
and \citeAY{Kozlov:1978:ARS}) applies. Then $\elt(\rst~',\rst)$ is independent of
the choice of $f(\rst_{0})$ (subject to it being smooth and only varying on the macroscopic scale), on the 
choice of $\phbo$, and (assuming statistical homogeneity) on the value of $\rst$. Then we may vary $f(\rst_{0})$
and $\phbo$ to change $<\ebvec>_{\Trst}$ and thus determine $\elt(\rst~',\rst)=\elt(\rst~')$ through \eq{Kdelt}.
For materials that are periodic inside $\GO$, with periodic cell much smaller than the size of $\GO$, $\elt(\rst~')$ can be obtained 
from the fields that solve the homogenization cell problem.%
\index{homogenization!cell problem}
i.e. with $\jb(\rst)$ and $\ebvec(\rst)$ having the same periodicity as the material, and the cell average of
$\ebvec(\rst)$ having any value we desire.

By assumption $<\ebvec>_{\Trst}$ has a smooth dependence on $\rst$
and so by taking the average of (\ref{Kdelt}) over points $\rst~'$ in the sphere $\Trst$
we can identify $\efl$ with the average of $\elt(\rst~')$,
\beq \efl=~<\elt>_{\Trst}= {\frac{1}{|\Trst|}}\int_{\Trst} d\rst~'~\elt(\rst~').\eeq{Kelrel}
To determine $\elrst$ and hence $\efl$ it suffices to prescribe 
linear potentials on the boundary $\partial\Omega$ of $\Omega$, i.e. to
suppose $f(\rst_{0})$ takes the form
\beq f(\rst_{0})= -\rst_{0} \cdot \vec{v}_{0}, \eeq{Klnbc}
where $\vec{v}_{0}$ is a constant vector.
Then the fields $\ebvec_{0}$ and $<\ebvec>_{\Trst}$ which solve the
constitutive equations in a homogeneous body are uniform,
\beq \ebvec_{0}=<\ebvec>_{\Trst}=\vec{v}_{0}\phbo . \eeq{Klinr}
Consequently for the purpose of determining both $\elt(\rst~')$ and $\efl$
the averages $<~~>_{\Trst}$ over each sphere ${\Trst}$
can be replaced by averages $<~~>_{\Omega}$ over the entire body $\Omega$.
Also to simplify subsequent formula let us select our dimensions
of length so that the body has unit volume,
\beq |\Omega|=1. \eeq{Kdiml}
Then, averages over $\Omega$ can be equated with integrals over $\Omega$.
From (\ref{Klinr}) and the relations (\ref{Kresp}) and (\ref{Kdelt}) 
of $\el(\rst~')$ and $\elt(\rst~')$ we have
\beq \el(\rst)= \elt(\rst)\vec{v}_{0}. \eeq{Klink}
where we have relabelled $\rst~'$ as $\rst$ to avoid confusion in the subsequent formulae. 
This, in conjunction with (\ref{Kconv}) and (\ref{Kelrel}), leads directly to the
expressions 
\beq \elrst=\int_{\Omega} d\rst~'~\srst, \eeq{Kels}
\beq \efl= \int_{\Omega} d\rst\int_{\Omega} d\rst~'~\srst, \eeq{Klas}
for the microscopic response tensor%
\index{response tensor!microscopic}
$\elrst$ and the effective tensor%
\index{effective tensor!formula}
$\efl$. Substitution of the series expansion for $\srst$ into these
expressions gives the desired series expansions%
\index{series expansions}
\beq \elrst=\lbo\onem+\sum_{s=1}^\infty\sum_{a_1,...,a_s=1}^p\sfac\Aa{a_1...a_s}
\epai{1}\lmo\epai{2}\lmo...\lmo\epai{s},  \eeq{Kselr}
\beq \efl=\lbo\onem+\sum_{s=1}^\infty\sum_{a_1,...,a_s=1}^p\sfac\als{a_1...a_s}
\epai{1}\lmo\epai{2}\lmo...\lmo\epai{s},  \eeq{Kself}
for $\elrst$ and $\efl$ in powers of the $\epa$'s with coefficients
\begin{eqnarray} 
\Aa{a_1...a_s}(\rst)=\int_{\Omega} d\rst~'~\kkam{a_1...a_s}(\rst,\rst~') ~~~~~~~~~~~~~~~~~~~~~~~~~~~~~
\nonumber \\
=\int_{\Omega} d\rst~'~[(I-\gam)\lama{1}\gam\lama{2}\gam...\gam\lama{s}]
(\rst,\rst~'), 
\label{Kcoaa}
\end{eqnarray}
\begin{eqnarray} 
\als{a_1...a_s}=\int_{\Omega} d\rst~\Aa{a_1...a_s}(\rst)~~~~~~~~~~~~~~~~~~~~~~~~~~~~~~~~~
\nonumber \\
=\int_{\Omega} d\rst\int_{\Omega} d\rst~'~[\lama{1}\gam\lama{2}\gam...\gam\lama{s}]
(\rst,\rst~'), 
\label{Kcoals}
\end{eqnarray}
where the prefactor of $(I-\gam)$ has been dropped from the last 
equation because $\gam$ acting upon any field produces a field with zero
integral over $\Omega$: see (\ref{Kinte}). 
\par
As a consequence 
of (\ref{Kproj}), (\ref{Kinte}) and (\ref{Klll}) the coefficients $\alps$ when $s>1$
satisfy 
\beq \sum_{a_1=1}^{p}\alps=0, \eeq{Ksuma}
\beq \sum_{a_s=1}^{p}\alps=0, \eeq{Ksumb}
\beq \sum_{a_i=1}^{p}\alps=
\als{a_1...a_{i-1}a_{i+1}...a_s}. \eeq{Ksumc}
In the special case $s=1$ (\ref{Klll}) implies
\beq \sum_{a=1}^{p}\als{a}=\onem. \eeq{Ksumdd}
Due to these identities it suffices, for any choice of reference 
index $q \in \{1,2,...p\}$,%
\index{reference index}
to consider the subset of coefficients
$\alps,s=1,2,....$ generated as the indices $a_i$ range over the 
reduced set $\{1,2,..q-1,q+1,...p\}$ skipping the reference index q.
The remaining coefficients $\alps$ where at least one index $a_i=q$
can then be recovered using (\ref{Ksuma})-(\ref{Ksumdd}).
In addition, recall from \eq{Kgsad4} that the operator $\gam$ is self-adjoint (this is also evident from \eq{Kdgamk} below).
Also ${\stackrel{\leftrightarrow}{\Lambda}}{}_{a}$ is obviously self-adjoint. So (\ref{Kcoals}) implies that the matrix $\alps$ is transformed
to its transpose under reversal
of the ordering of its subscripts: 
\beq \als{a_s a_{s-1} ... a_2 a_1}=(\als{a_1 a_2 ...a_{s-1} a_s})^{T}. \eeq{Krev}
\par
There are further identities satisfied by the coefficients $\alps$. In
particular, the first order coefficients satisfy
\beq \Tr(\als{a})=\Tr(\alss{l\bet})=\int_{\Omega} d\rst~\Tr(\lalb)=m_{l} f_{l},
\eeq{Ktra}
where $f_{l}$ denotes the volume fraction occupied by component $l$ and
$m_l$ takes values 1,2 or 3 according to whether the component $l$ has
orthorhombic symmetry,%
\index{orthorhombic symmetry}
uniaxial symmetry,%
\index{uniaxial symmetry}
or isotropic symmetry.%
\index{isotropic symmetry}
The last identity in \eq{Ktra} follows immediately for orthorhombic components by taking the trace in
\eq{Kccc} (i.e. $\Tr[\lalb(\rst)]=\xlr$), and for isotropic components by taking the trace in \eq{Kddd} (i.e. $\Tr[\lalb(\rst)]=3\xlr$).

The trace of the second order
coefficient $\als{ab}$ can also be easily evaluated when the components are
isotropic. To see this let us, for simplicity, suppose that the composite
material is periodic with periodicity $h$ much smaller than the dimensions
of $\Omega$. The action of $\gam$ on any $h$-periodic vector 
field $\vec{u}(\rst~')$ is local in Fourier space and produces a vector
field $\vec{v}(\rst)$ given by (\ref{Kdgam}) with Fourier components
\beq \vec{v}(\vec{k})=\gam(\vec{k})\vec{u}(\vec{k}), \eeq{Kgloc}
in which $\vec{u}(\vec{k})$ denotes the Fourier component 
of $\vec{u}(\rst)$ and where the matrix $\gam(\vec{k})$ has elements
\begin{eqnarray}
\{\Gamma_1\}_{ij} (\vec{k})=k_{i} k_{j}/|\vec{k}|^{2} ~~~~~~ \vec{k} \ne 0,
\nonumber \\
= 0 ~~~~~~~~~~~~~~~ \vec{k}=0.
\label{Kdgamk}
\end{eqnarray}
Clearly (\ref{Kdgamk}) implies 
\begin{eqnarray}
\Tr(\gam(\vec{k}))=1 ~~~~ \vec{k} \ne 0,
\nonumber \\
= 0 ~~~ \vec{k}=0,
\label{Ktrgam}
\end{eqnarray}
and it follows that the operator
\beq \Gamma(\rst,\rst~') \equiv \Tr(\gam(\rst,\rst~')) \eeq{Kdtgam}
acts on any $h$-periodic scalar field $u(\rst)$ to produce the scalar field
\beq v(\rst)=\int_{\Omega} d\rst~'~\Gamma(\rst,\rst~'~)u(\rst)
=u(\rst)~-~\int_{\Omega} d\rst~'~u(\rst~'). \eeq{Ktgamr}
When the components are isotropic $\lama{}(\rst)=\Lambda_{a}(\rst)\onem$ and we have
\begin{eqnarray}
\Tr(\als{ab})=\int_{\Omega} d\rst \int\ d\rst~'~[\Lambda_{a}\Gamma\Lambda_{b}]
(\rst,\rst~'~)~~~~~~~~~~~~~~~~~~~~~~~
\nonumber \\
=\int_{\Omega} d\rst\Lambda_{a}(\rst)\Lambda_{b}(\rst)
-\int_{\Omega} d\rst\Lambda_{a}(\rst)~\int_{\Omega} d\rst~'\Lambda_{b}(\rst~')
\nonumber \\
=\delta_{ab}f_{a}~-~f_{a} f_{b},~~~~~~~~~~~~~~~~~~~~~~~~~~~~~~~~~~~~~~~~~
\label{Ksecre}
\end{eqnarray}
where again $f_{a}$ and $f_{b}$ are the volume fractions of the components $a$ and $b$.
\par
In two-dimensional composites (\ref{Ksecre}) is a simple corollary of 
one of an 
infinite set of identities satisfied by the coefficients $\alps$. These
follow from the simple duality%
\index{duality relations}
observation (see, for example, \citeAPY{Keller:1964:TCC},
\citeAPY{Dykhne:1970:CTD} and \citeAPY{Mendelson:1975:TEC}) that a $90^{\circ}$ rotation, $\rprperp$ acting
on a curl-free field%
\index{curl-free fields}
produces a divergence-free field%
\index{divergence-free fluxes}
and vice versa. Equivalently, from (\ref{Kdgamk}) we see immediately that
\beq \rprperp\gam(\rprperp)^{T}=I-\gam-\gamo, \eeq{Kacrp}
or alternatively,
\beq \rprperp\gam=(I-\gam-\gamo)\rprperp, \eeq{Kacr}
where $\gamo(\rst,\rst~')$
is the  operator which simply acts to average the field:
 $\gamo(\rst,\rst~')$ acting on a field $\vec{u}(\rst~')$ produces
the uniform field
\beq \vec{v}(\rst)= \int_{\Omega}d\rst~'~\gamo(\rst,\rst~')\vec{u}(\rst~')
=\int_{\Omega}d\rst~'~\vec{u}(\rst~'), \eeq{Kdefgo}
and $\rprperp$ is the operator which acts locally upon a 
field $\vec{u}(\rst)$ rotating it by $90^{\circ}$ to produce the 
field $\vec{v}(\rst)$ with elements
\beq v_{\alpha}=\sum_{\beta=1}^{2}R^{\bot}_{\alpha \beta}u_\beta, \eeq{Ktwrot}
where $R^{\bot}_{\alpha \beta}$ are in turn the elements of the matrix
\begin{eqnarray}  
\BR_\bot=\left[ \begin{array}{cc} ~0 & ~1 \\ -1 & ~0 \end{array}
 \right],  
\label{Kdefrp}
\end{eqnarray}
for a  $90^{\circ}$ rotation.
Accordingly we can use (\ref{Kacrp}) to express $\rprperp\alps(\rprperp)^{T}$
as a linear combination of the coefficients $\als{a_1...a_m}$ with
$m \le s$. For example, if the components are isotropic $\rprperp$
commutes with $\lama{}$ and we have
\begin{eqnarray}
\rpj{}\als{a_1 a_2}(\rpj{})^{T}=\int_{\Omega} d\rst\int_{\Omega} d\rst~'
~[\lama{1}(I-\gam-\gamo)\lama{2}](\rst,\rst~')
\nonumber \\
=-\als{a_1 a_2}+\del_{a_1 a_2} f_{a_1}\onem-f_{a_1}f_{a_2}\onem,~~~~~~~~~~~~
\label{Kterma}
\end{eqnarray}
\begin{eqnarray}
\rpj{}\als{a_1 a_2 a_3}(\rpj{})^{T}=
\int_{\Omega} d\rst\int_{\Omega} d\rst~'
~[\lama{1}(I-\gam-\gamo)\lama{2}(I-\gam-\gamo)\lama{3}](\rst,\rst~')
\nonumber \\
=\als{a_1 a_2 a_3}-\del_{a_1 a_2}\als{a_2 a_3}-\als{a_1 a_2}\del_{a_2 a_3}
+f_{a_1}\als{a_2 a_3}+\als{a_1 a_2}f_{a_3}~~
\nonumber \\
+\del_{a_1 a_2}\del_{a_2 a_3}f_{a_1}\onem
-\del_{a_1 a_2}f_{a_2}f_{a_3}\onem-f_{a_1}f_{a_2}\del_{a_2 a_3}\onem
+f_{a_1}f_{a_2}f_{a_3}\onem.
\nonumber \\
~
\label{Ktermb}
\end{eqnarray}
The identity (\ref{Ksecre}) is easily seen to follow from (\ref{Kterma}) by
taking the trace of that equation. We now return to considering three
dimensional composite materials.
\par
When only one field is present, i.e. $n=1$, then the knowledge of
the series expansion of $\efl$ in powers of the $\epa$ up to
a given order $s$ is insufficient to determine the coefficients
$\alps$ when $p \ge 3$ and $s \ge 3$. For example, consider the
problem of electrical conductivity,
\beq \diver J(\rst)=0,~~\grad\times E(\rst)=0,~~J(\rst)=\si(\rst)E(\rst),
~~\si(\rst)=\sum_{a=1}^{p}\sia{a}\lama{}, \eeq{Kcond}
in a nearly homogeneous, nearly isotropic 
material with small values of the conductivity differences
\beq \vepa{}=\sia{a}-\sia{o}. \eeq{Kcodif}
Since the scalar quantities $\sia{o}$ and $\vepa{}$ commute,
(\ref{Kself}) reduces to the well-known series expansion%
\index{series expansions!conductivity}
for the
effective conductivity 
\beq \sigs=\sia{o}\onem+\sum_{s=1}^\infty\sum_{a_1,...,a_s=1}^p\sfac\bls{a_1...a_s}
\vepa{1}\vepa{2}...\vepa{s}/(\sia{o})^{s-1}, \eeq{Kcoexp}
with coefficients
\beq \blps=[{{\stackrel{\leftrightarrow}{\alpha}_{a_1...a_s}}}]_{\rm{sym}}
\equiv \frac{1}{s!}\sum_{\rm{\small{permutations}}}{{\stackrel{\leftrightarrow}{\alpha}_{p(a_1...a_s)}}}, \eeq{Ksymco}
where the brackets $[~~]_{\rm{sym}}$ denote a symmetrization%
\index{symmetrization!indices}
over all $s!$ permutations $p(a_1...a_s)$ of the field indices $a_1...a_s$,
excluding the space indices. In view of (\ref{Krev}) we have, for example,
\begin{eqnarray}
\bls{a_1}=\als{a_1},~~~~
\bls{a_1 a_2}=\frac{1}{2}[ \als{a_1 a_2}+(\als{a_1 a_2})^{T}],~~~~~~~~~~~~~~~~~~~
\nonumber \\
\bls{a_1 a_2 a_3}=\frac{1}{6}[ \als{a_1 a_2 a_3}+ \als{a_2 a_3 a_1}+ \als{a_3 a_1 a_2}+( \als{a_1 a_2 a_3}+\als{a_2 a_3 a_1}+\als{a_3 a_1 a_2})^{T}].
\nonumber \\
~
\label{Krealbe}
\end{eqnarray}
If the coefficients $\bls{a_1...a_j}$ are known for all $j \le m$ 
then it is clearly impossible to recover all the coefficients 
$\alps$ for $s \le m$: one can only recover the linear combinations
given by (\ref{Ksymco}). However this does not eliminate the
possibility that the coefficients $\alps$ could
be recovered from knowledge of the entire infinite set of
coefficients $\bls{a_1...a_j}$. As we will see in the Section 7
the value $\alps$ can take is nonlinearly correlated with the coefficients
$\bls{a_1...a_j}$ with $j \le m$ through a set of matrix inequalities
and it is conceivable that these matrix inequalities are sufficiently
stringent to uniquely determine a given coefficient $\alps$ as $m$
tends to infinity.
\section[The weights and normalization matrices]{The weights and normalization matrices and a stratification of the Hilbert space }
\setcounter{equation}{0}

Suppose the coefficients $\Aa{a_1...a_s}(\rst)$ of the 
microscopic response tensor $\elrst$ are known as functions of $\rst$, 
for all $s$ up to a given
order $m$, and for all combinations of indices $a_i$ taken from the
set $\{1,2,...,p\}$. In light of (\ref{Kcoals}) 
one might think that this information
would only be sufficient to determine the coefficients $\alps$ 
for $s\leq m$.
However this does not take into account the relations
\begin{eqnarray} 
\int_{\Omega} d\rst~ [\Aa{a_i a_{i-1}...a_1}(\rst)]^{T}\Aa{a_{i+1} a_{i+2}...a_s}(\rst) ~~~~~~~~~~~~~~~~~~~~~~~~~~~~~~~~~~~~~~~~
\nonumber \\
= \int_{\Omega} d\rst \int_{\Omega} d\rst~'~[
\lama{1}\gam...\lama{i}(I-\gam)\lama{i+1}\gam...\lama{s}](\rst,\rst~') 
\nonumber \\ 
= \delta_{a_i a_{i+1}}\als{a_1 a_2 ...a_{i-1} a_{i+1}... a_s}-\alps,~~~~~~~~~~~~~~~~~~~~~~~~~~ 
\label{Kprod}
\end{eqnarray}
implied by (\ref{Kcoaa}), (\ref{Kproj}) and (\ref{Keee}).
These relations, which hold for all $i \in \{1,2,...,s-1\}$, 
allow the coefficients $\alps$ to be
determined for $s\leq 2m$ from knowledge of the functions $\As(\rst)$ for
all $s\leq m$.
\par
Now note that (\ref{Kcoals}) 
and (\ref{Kprod}) imply inequalities such as the 
positive semidefiniteness of the tensors $\als{a a}$ 
and $\als{a}-\als{a a}$ , $a=1,2...p$. The question of
what other inequalities apply to the coefficients $\alps$ has been
analyzed in depth by Milton (\citeyearNP{Milton:1987:MCEa}, \citeyearNP{Milton:1987:MCEb}). Briefly, and as
proved later section 7, the set of
coefficients $\alps$ for $s \leq 2m$ derive from, and in turn uniquely 
determine,  a set of normalization 
matrices%
\index{normalization matrices}
$\norj$, $j=1,2,...m$, and weight 
matrices%
\index{weight matrices}
$\wej{a}$, $a=1,2,...p$, $j=0, 1,2,...m-1$
that are real and symmetric and satisfy
\beq \norj \geq 0,~~~~\wej{a} \geq 0,~~~~
\sum_{a=1}^p\wej{a}=\onej,
\eeq{Knaw}
where $\onej$ denotes the $k$-dimensional identity matrix, where in a
space of $3$ dimensions, $k=3(p-1)^{j}$.
These matrices have elements $N^{j}_{\tau ,\mu}$, 
$W^{j}_{a ,\tau ,\mu}$ and
\beq I^{j}_{\tau ,\mu}=\delta_{\tau \mu}, \eeq{Kide}
labeled by strings $\tau=a_1 a_2 ...a_j \alp$ and
$\mu=b_1 b_2...b_j \bet$ of integers $a_i$ or $b_i,~i=1,2,..j$ 
chosen from the set $\{ 1,2,...q-1,q+1...p\}$ (skipping the reference
index%
\index{reference index}
$q$)
terminated by a single space index $\alp$ or $\bet$ chosen from
the set $\{ 1,2,3\}$. Thus each matrix has dimension 
$3(p-1)^{j}$ dependent  on $j$, for $p>2$.

\par
Conversely, if a set of $\alps$ derive from any sequence 
of $3(p-1)^{j}$-dimensional symmetric real matrices $\norj, j=1,2,..$ 
and $\wej{a}, a=1,2,...p$,$~j=0,1,2,...$ satisfying (\ref{Knaw}) then
there always exists a set of commuting 
projection operators, $\lama{}, a=1,2,..p$, satisfying (\ref{Keee}) and
 (\ref{Klll}), and another noncommuting projection operator $\gam$, 
satisfying (\ref{Kproj}) 
such that $\alps$ is given by (\ref{Kcoals}): we will see in Section 7 that, with a suitable choice of basis, the 
operators $\lama{}$ only depend on the weight matrices, 
while $\gam$ in this representation only depends
on the normalization matrices. However not every sequence of 
normalization and weight
matrices corresponds to a composite: there are additional subtle
restrictions on the operators $\lama{}$ and $\gam$ in a composite
which lead to nontrivial restrictions on the coefficients $\alps$.
In particular, as noticed by \citeAPY{Zhikov:1994:HDO}, when all the components are isotropic
a theorem of \citeAPY{Meyers:1963:EGS}%
\index{Meyers theorem}
implies that, in the
limit as the volume fraction $f_a$ of component $a$ tends to zero, $\efl$
cannot depend on $\lba$ unless of course $\lba$ has infinite
or zero eigenvalues. In other words there exist inequalities
which force any coefficient $\alps$,
with $a_i=a$ for some $i \in \{1,2,...s\}$, to approach zero 
as $f_{a}=\Tr(\als{a})$ tends to zero.
\par
It remains to link the expansion coefficients with the weight and
normalization matrices and to derive suitable representations for
the operators $\lama{}$ and $\gam$.
In the rest of the
Chapter, lower-case greek letters, other 
than $\alp$ or $\bet$ will always
be used to denote strings of indices,
where each index except the last is an element of the set
 $\{ 1,2,...q-1,q+1...p\}$ and where the final space index takes 
values from the set $\{ 1,2,3\}$. The length
$j$ of a string will refer to the number of indices in the string
excluding the final space index.  Also we use commas to
separate strings of indices that label the elements of a matrix. 
Finally, a $\leftrightarrow$ above a character accompanied by
a superscript $j$ will indicate a $3(p-1)^{j}$ dimensional 
matrix in the string indices,with strings of length $j$.
\par
First consider the sequence of fields obtained in the following
fashion. We begin with a set of three or two uniform fields $\xej_{\alp}$,
$(\alp=1,2,3)$ each aligned with its corresponding coordinate axis.
[The notation is somewhat bad as $\xej_{\alp}$ should not be confused with
a variable or spatial coordinate, but it follows the notation given 
in appendix 1 of \citeAPY{Milton:1987:MCEa}.]
Then we set
\beq \pej_{a_1 a_2 ...a_k \alp}(\rst)=\lama{1}\gam\lama{2}\gam...
 \gam\lama{k}\xej_{\alp}, \eeq{Kdpvec} 
\beq \eej_{a_1 a_2 ...a_k \alp}(\rst)=\gam\lama{1}\gam\lama{2}\gam...
 \gam\lama{k}\xej_{\alp}. \eeq{Kdevec}
Note that the response 
coefficients%
\index{response coefficients}
$\Aa{a_1...a_s}(\rst)$  derive from these fields:
from (\ref{Kcoaa}) we have
\beq \Aa{a_1...a_s}(\rst)\xej_{\alp}=
\pej_{a_1 ...a_s \alp}(\rst)-\eej_{a_1 ...a_s \alp}(\rst).
\eeq{Krape}
Introducing the standard inner product,%
\index{inner product}
\beq (\vec{u},\vec{v})=\int_\Omega d\rst ~\overline{\vec{u}(\rst)} \cdot \vec{v}(\rst), 
\eeq{Kdinpr}
between any two real fields $\vec{u}(\rst)$ and $\vec{v}(\rst)$, where the overline
denotes complex conjugation, it is 
clear (see also (\ref{Kprod})) that the inner product between any
pair of the above fields can be written in terms of the elements
of the coefficient matrix $\alps$: we have
\beq (\eej_{\tau},\eej_{\eta})=\alpha_{\bar{\tau} \eta},\eeq{Kinpa}
\beq (\eej_{\tau},\pej_{\eta})=(\pej_{\tau},\eej_{\eta})
=\alpha_{\bar{\tau} \eta},\eeq{Kinpb}
\beq (\pej_{a \tau},\pej_{b \eta})
=\delta_{ab}\alpha_{\bar{\tau} a \eta},
\eeq{Kinpc}
where $\tau$ and $\eta$ represent strings of indices of lengths $j$ and
$k$ respectively, and $\bar{\tau}$ is obtained from $\tau$ by reversing
the sequence of indices in the string.
\par
The space spanned by these fields has a natural stratification%
\index{subspace collections!stratification}
into a sequence of orthogonal subspaces $\Xc{0},\Yc{1},\Xc{1},\Yc{2},
\Xc{2},...$ The subspace $\Xc{0}$ is defined as the subspace spanned
by the uniform fields $\xej_{\alpha},~\alpha=1,2,3$. Let $\cfj{j}$ denote
the subspace spanned by the fields $\xej_{\alpha}, \pej_{\eta}(\rst)$ 
and $\eej_{\eta}(\rst)$ as $\eta$ ranges over all strings of length
$j$. Also let $\cgj{j}$ denote the closure of $\cfj{j-1}$ under the action
of the set of operators $\lama{}~a=1,2,..p$: this is the space spanned
by $\cfj{j-1}$ and fields $\pej_{\tau}$ as $\tau$ ranges over strings of
length $j$. Note that $\cfj{j}$ in turn is the closure of $\cgj{j}$ 
under the action of $\gam$. These subspaces satisfy the 
inclusion relations
\beq \Xc{0}=\cfj{0}\subset\cgj{1}\subset\cfj{1}\subset\cgj{2}
\subset\cfj{2}\subset\cgj{3}\cdots.\eeq{Kinclu}
Accordingly we define $\Yc{j}, j=1,2,...$ as the subspace of $\cgj{j}$
which is the orthogonal complement of $\cfj{j-1}$, and $\Xc{j}, j=1,2,...$ 
as the subspace of $\cfj{j}$ which is the orthogonal 
complement of $\cgj{j}$. 
\par
The weights and normalization matrices%
\index{weight matrices}%
\index{normalization matrices}%
are obtained through the 
introduction of an orthonormal basis set of fields, comprised of fields
$\xej_{\eta}(\rst)$, denoted as type $x$, and fields $\yyej{\eta}(\rst)$, 
denoted as type $y$,
generated by a special version of Gram-Schmidt
orthogonalization%
\index{Gram-Schmidt orthogonalization}
applied to the sequence of 
fields $\pej_{\tau}(\rst)$ and $\eej_{\tau}(\rst)$. These basis
fields $\xej_{\eta}(\rst)$ and $\yyej{\eta}(\rst)$ 
will be called fields of order $j$ if the
string $\eta$ has length $j$. Any linear combination of type $x$
(or type $y$) basis fields of order $j$ will also be called
a type $x$ (or type $y$) field of order $j$ and we will establish
that these type $x$ (or type $y$) fields of order $j$ are 
precisely the fields in the subspace $\Xc{j}$ (or $\Yc{j}$).

\section[Construction of the basis fields]{Construction of the basis fields and weights and normalization factors}
\setcounter{equation}{0}

 Those readers not interested in
the details of the construction of the basis fields%
\index{basis fields}
and weight
and normalization matrices can skip to Section 9. We follow the
construction procedure outlined in Appendix 1 of \citeAPY{Milton:1987:MCEa}. Recall 
that the uniform fields $\xej_{\alp}$ are already defined. 
Let us therefore suppose, for some $j \ge 1$, that all
type $x$ basis fields of order $j-1$ have been introduced.
The weight matrices $\wejl{a}$%
\index{weight matrices}
are then defined via
\beq \wjl_{a,\omega,\rho}
\equiv (\xej_{\omega},\lama{}\xej_{\rho}), \eeq{Kweit}
where $\omega$ and $\rho$ are strings of length $j-1$.
Next we introduce the first set of auxiliary fields
\beq \aej{a\omega}(\rst) \equiv \lar\xej_{\omega}(\rst)
-\sum_{\zeta}^{}\wjl_{a,\omega,\zeta}\xej_{\zeta}(\rst), \eeq{Kdefna}
which are defined in this way to ensure orthogonality to the previous set
of type $x$ fields of order $j-1$. Also from (\ref{Klll}) it is evident
that
\beq \sum_{a=1}^{p} \aej{a\omega}(\rst)=0, \eeq{Ksumn}
and consequently it suffices to consider the subset of fields
$\aej{a\omega}(\rst)$ as the index $a$ ranges over the reduced set
$\{1,2,..,q-1,q+1,...p\}$. The inner products between the fields
in this subset are given by
\beq (\aej{a\omega},\aej{b\rho})=\yj_{a\omega,b\rho}, \eeq{Kinna}
where
\beq \yj_{a\omega,b\rho} \equiv \delta_{ab}\wjl_{a,\omega,\rho}
-\sum_{\zeta}\wjl_{a,\omega,\zeta}\wjl_{b,\zeta,\rho}, \eeq{Kydefs}
and the indices $a$ and $b$ belong to the reduced set (as does any
other index in the strings $\omega$ and $\rho$ apart from the
terminating index).
We normalize these fields to obtain the desired family of type $y$ 
basis fields of order $j$,
\begin{eqnarray}
\yyej{b\rho} \equiv \sum_{a\ne q} \sum_{\omega} C^{j}_{b\rho,a\omega}
\aej{a\omega} ~~~~~~~~~~~~~~~~~~~~~~
\nonumber \\
=\sum_{a\ne q} \sum_{\omega} C^{j}_{b\rho,a\omega}
(\lama{}\xej_{\omega}
-\sum_{\zeta}^{}\wjl_{a,\omega,\zeta}\xej_{\zeta}),
\label{Kdefyf}
\end{eqnarray}
where 
\beq \cej \equiv (\yej)^{-1/2}. \eeq{Kdefcc}
Similarly, starting from these fields, let us introduce the commuting pair
of matrices
\beq U^{j}_{\tau,\phi} \equiv (\yyej{\tau},\gam\yyej{\phi}), \eeq{Kdefu}

\beq V^{j}_{\tau,\phi} \equiv (\yyej{\tau},(I-\gam)\yyej{\phi})
=\delta_{\tau\phi}-U^{j}_{\tau,\phi}, \eeq{Kdefv}
where the string indices $\tau$ and $\phi$ are now of length $j$.
In terms of these matrices the normalization matrix%
\index{normalization matrices}
is defined via
\beq \norj \equiv (\uej)^{-1}-\onej, \eeq{Krenu}
implying
\beq \uej=(\onej+\norj)^{-1},~~~~~~~~\vej=\{(\onej+(\norj)^{-1}\}^{-1}.
\eeq{Kuvan}
Next we generate the second set of auxiliary fields
\beq \bej{\tau}(\rst) \equiv \int_{\Omega} d\rst~'~\gam(\rst,\rst~')\yyej{\tau}
(\rst~')~ -~\sum_{\nu} U^{j}_{\tau,\nu}\yyej{\nu}(\rst), \eeq{Kdefib}
which are orthogonal to the fields $\yyej{\phi}$, and have inner products%
\index{inner product}
\beq (\bej{\tau},\bej{\phi})=\sum_{\nu}^{} 
U^{j}_{\tau,\nu}V^{j}_{\nu,\phi}.
\eeq{Kinnb}
Normalizing these fields then produces the next 
orthonormal set of type $x$ basis fields of order $j$:
\beq
\xej_{\phi} \equiv \sum_{\tau}D^{j}_{\phi,\tau}\bej{\tau}
=\sum_{\tau}D^{j}_{\phi,\tau}(\gam\yyej{\tau}
~ -~\sum_{\nu} U^{j}_{\tau,\nu}\yyej{\nu}),
\eeq{Kdnexx}
where 
\beq \dej \equiv (\uej\vej)^{-1/2}
=(\norj)^{1/2}+(\norj)^{-1/2}. \eeq{Kdefdd}
By induction this completes the definition of the basis fields, and
weight and normalization matrices.

\par
From the definitions (\ref{Kweit}),(\ref{Kydefs}),
 (\ref{Kdefu}) and (\ref{Kdefv}) it is
clear that the matrices $\wejl{a},\yej,\uej$ and $\vej$ are positive
semidefinite. Furthermore from (\ref{Kuvan}) and from the orthonormality
of the sets of fields, $\xej_{\omega}$ and $\yyej{\tau}$
 it follows that the weights and
normalization matrices satisfy (\ref{Knaw}).
 We avoid considering the rather special limiting 
case where the matrices  $\wejl{a},\uej$ and $\vej$
have zero eigenvalues. In this event the matrices $\yej$
and $\uej\vej$ become singular and technical difficulties  
arise in the above construction procedure because the inverses
needed in (\ref{Kdefcc}) and (\ref{Kdefdd}) do not exist.
\par
The set of normalization and weight matrices obtained in this
way clearly depend on the choice of reference component $q$.
However the subspace spanned by type $x$ (or type $y$)
fields of order $j$ remains invariant: it is only the basis within
each subspace that changes when the choice of reference 
component is changed. Consequently the eigenvalues of the weight
and normalization matrices do not depend on the choice of
reference media.
\par 
Observe from (\ref{Kdefyf}) and (\ref{Kdnexx}) that for $a \ne q$
\beq \lama{}\xej_{\omega}=\sum_{\zeta}\wjl_{a,\omega,\zeta}\xej_{\zeta}
+\sum_{b\ne q}\sum_{\rho}\mj_{a\omega, b\rho}\yyej{b\rho}, \eeq{Kactl}
\beq \gam\yyej{\tau}=\sum_{\nu}U^{j}_{\tau,\nu}\yyej{\nu}
+\sum_{\phi}X^{j}_{\tau,\phi}\xej_{\phi}, \eeq{Kactg}
where 
\beq \cxej \equiv (\uej\vej)^{1/2}=\{(\norj)^{1/2}+(\norj)^{-1/2}\}^{-1}. 
 \eeq{Kdefcx}

\beq \mej{} \equiv (\yej)^{1/2}=(\cej)^{-1}, \eeq{Krooty} 
and $\yej$ in turn is given by (\ref{Kydefs}). 
\par
Applying $\lamc$, with $c \ne q$ 
to both sides of this first equation and $\gam$ to
both sides of the second equation gives
\beq \sum_{b\ne q}\sum_{\rho}\mj_{a\omega, b\rho}\lamc\yyej{b\rho}
=\sum_{\zeta}(\delta_{ac}\delta_{\omega\zeta}-
\wjl_{a,\omega,\zeta})\lamc\xej_{\zeta}, \eeq{Ksacy}
\beq \sum_{\phi}X^{j}_{\tau,\phi}\gam\xej_{\phi}
=\sum_{\nu}V^{j}_{\tau,\nu}\gam\yyej{\nu}. \eeq{Ksacg}
Substituting (\ref{Kactl}) and (\ref{Kactg}) back into these expressions
produces after some algebraic manipulation,
\beq \lamc\yyej{b\rho}
=\sum_{a\ne q}\sum_{\zeta}Q^{j}_{c,b\rho,a\zeta}\yyej{a\zeta}
+\sum_{\zeta}\mj_{b\rho, c\zeta}\xej_{\zeta}, \eeq{Kacly}
\beq \gam\xej_{\nu}=\sum_{\phi}V^{j}_{\nu,\phi}\xej_{\phi}
+\sum_{\phi}X^{j}_{\nu,\phi}\yyej{\phi}, \eeq{Kacgx}
where $\qej{c}$ is the matrix,
\beq  \qej{c} \equiv \mej{c}(\wejl{c})^{-1}(\mej{c})^{T}, \eeq{Kdefq}
and $\mej{a}$, with transpose $(\mej{a})^{T}$, is the rectangular 
submatrix of the square matrix $\mej{}$ defined in (\ref{Krooty})
with elements $\mj_{a\tau,\lambda}$ labeled by
the strings $\tau$ and $\lambda$.
\par
So $\lamc$ acting upon any basis field
produces a linear combination of two fields: one field of the
same order and type as the basis field and the other field
of adjacent order and opposite type. By contrast
$\gam$ acting on any basis field  produces 
a field of the same order but mixed type.
\par
By construction the basis fields of a given order
form an orthonormal set.
To establish the orthonormality of the entire basis set we still 
need to show that the basis fields of order $j$ are orthogonal
to the subspace spanned by the fields
of order at most $j-1$. Note that this subspace can also be identified
with the subspace $\cfj{j-1}$ spanned by the fields $x_\alpha$,
$\pej_{\eta}(\rst)$, and $\eej_{\eta}(\rst)$ as $\eta$ ranges over
strings of length $k \le j-1$.
We argue by induction and begin by assuming that the collection of fields
$\xej_{\eta}$, and $\yyej{\eta}$ of order at most $j-1$
forms an orthonormal basis of $\cfj{j-1}$: this
is clearly true when $j=1$ because then ${\cal{F}}^{0}$ is the 
 three dimensional
space spanned by the fields $\xej_{\alpha}$. In particular
the assumption implies that
within $\cfj{j-1}$ basis fields of different types
or different orders are orthogonal. Since $\lama{}$ is
self-adjoint (\ref{Kdefna}) implies that
\beq (\aej{a\omega},\xej_{\eta})=(\xej_{\omega},\lama{}\xej_{\eta}),~~~~~
(\aej{a\omega},\yyej{\eta})=(\xej_{\omega},\lama{}\yyej{\eta}),
 \eeq{Kipaxy}
where the string  $\omega$ has length $j-1$.
The choice of auxiliary fields guarantees that the first inner product
is zero when the length $k$ of the string $\eta$ equals $j-1$. It is
also zero when $k<j-1$ because then (\ref{Kactl}) 
implies $\lama{}\xej_{\eta} \in \cgj{j-1}$ where $\cgj{j-1}$ can now 
be identified with the space spanned by fields 
in $\cfj{j-2}$ and type $y$ fields of order $j-1$. 
Similarly the second inner product is zero because (\ref{Kacly}) 
implies $\lama{}\yyej{\eta} \in \cgj{j-1}$. Since these
inner products are zero we conclude
that the auxiliary fields $\aej{a\omega}$ are orthogonal to $\cfj{j-1}$.
The type $y$ fields of order $j$ are linear combinations of these 
auxiliary fields and so must also be orthogonal to the space $\cfj{j-1}$. 
Analogous considerations show that the inner products
\beq (\bej{\tau},\xej_{\eta})=(\yyej{\tau},\gam\xej_{\eta}),~~~~~
(\bej{\tau},\yyej{\eta})=(\yyej{\tau},\gam\yyej{\eta}), \eeq{Kipbxy}
implied by (\ref{Kdefib}) are zero when  the string $\tau$ has length $j$.
We deduce that the type $x$ fields of order $j$ are also orthogonal to the 
space $\cfj{j-1}$. This completes the proof of orthonormality of the basis.
As a corollary, it follows that $\Xc{j}$ and $\Yc{j}$ 
represent respectively the  type $x$ fields and type $y$ fields 
of order $j$.
\section[Representation of the projection operators]{Representation of the projection operators and recovery of weight and normalization matrices
from series expansion coefficients}
\labsect{Krepop}
\setcounter{equation}{0}

	Clearly (\ref{Kactl}) and (\ref{Kacly}) determine the action of
$\lama{}$ on the basis fields while (\ref{Kactg}) and (\ref{Kacgx}) determine
the action of $\gam$. It immediately follows that the projection operators%
\index{projection operators!representation}%
\index{representation!projection operators}
$\gam$ and $\lama{}$ 
for $a \ne q$ are represented in this basis by the block tridiagonal 
infinite matrices%
\index{block tridiagonal matrices}
\begin{eqnarray}
\lama{} & = &\left[ \begin{array}{ccc} \begin{array}{rr} \wea{0}~~ & \mea{1}\\
(\mea{1})^{T} & \qea{1} \end{array} & 0 & \\ 0 & \begin{array}{rr}
\wea{1}~~ & \mea{2} \\ (\mea{2})^{T}& \qea{2} \end{array} & \\  &  & \ddots \end{array}
\right], \nonumber \\
\gam & = & \left[ \begin{array}{cccc} 0 & 0 & 0 &\\ 0 & \begin{array}{rr}
\ue{1} & \xe{1} \\ \xe{1} & \ve{1} \end{array} & 0 & \\ 0 & 0 &
\begin{array}{rr} \ue{2} & \xe{2} \\ \xe{2} & \ve{2} \end{array}
& \\ & & & \ddots \end{array} \right].
\nonumber \\
\label{Kbigm}
\end{eqnarray}
The blocks in these matrices act upon fields of the order indicated by the
block superscript, with the exception of the rectangular 
blocks $(\mej{a})^{T}$ which act on fields of order $j-1$. The blocks
going across a given row act on fields alternating
between type $x$ and type $y$,
beginning with type $x$. The tridiagonal form%
\index{tridiagonal form}
of the matrices
representing  $\gam$ and $\lama{}$   reflects
the fact that the procedure for constructing the basis fields is
similar to the procedure used in the Lanczos algorithm%
\index{Lanczos tridiagonalization algorithm}
for tridiagonalization of 
symmetric matrices (see, for example, \citeAY{Strang:1986:IAM2}). The operator
\beq \lamq=I-\sum_{a \ne q} \lama{} \eeq{Kdeflq}
also can be represented by the matrix in (\ref{Kbigm}) with $a=q$
provided we define $\mej{q}$ via 
\beq \mej{q} \equiv -\sum_{a \ne q} \mej{a}, \eeq{Kdefmq}
and $\qej{q}$ via (\ref{Kdefq}). 
\par
The matrix representing $\lama{1}\gam\lama{2}\gam...\gam\lama{s}$
is generated by taking products of the matrices in (\ref{Kbigm}).
Also for any operator $\brst$ 
with elements $B_{\alpha\beta}(\rst,\rst~')$ we have
\beq \int_{\Omega}d\rst \int_{\Omega}d\rst~'~B_{\alpha\beta}(\rst,\rst~')
=(\vec{x}_{\alpha},\brt\vec{x}_{\beta}). \eeq{Kcores}
In particular then $\alps$ is the first block which appears in
the matrix representing $\lama{1}\gam\lama{2}\gam...\gam\lama{s}$.
In this way we obtain expressions, such as
 \beq \als{a_1}=\wejz{a_1}, \eeq{Kreaa}
\beq \als{a_1 a_2}=\mejo{a_1}\ue{1}(\mejo{a_2})^{T}, \eeq{Krebb}
\beq \als{a_1 a_2 a_3}=\mejo{a_1}\ue{1}\qejo{a_2}\ue{1}(\mejo{a_3})^{T}
 +\mejo{a_1}\xe{1}\wejo{a_2}\xe{1}(\mejo{a_3})^{T}, \eeq{Krecc}
\begin{eqnarray}
\als{a_1 a_2 a_3 a_4}
=\mejo{a_1}\ue{1}\qejo{a_2}\ue{1}\qejo{a_3}\ue{1}(\mejo{a_4})^{T}
+\mejo{a_1}\ue{1}\qejo{a_2}\xe{1}\wejo{a_3}\xe{1}(\mejo{a_4})^{T}
\nonumber \\
+\mejo{a_1}\xe{1}\wejo{a_2}\xe{1}\qejo{a_3}\ue{1}(\mejo{a_4})^{T}
+\mejo{a_1}\xe{1}\wejo{a_2}\ve{1}\wejo{a_3}\xe{1}(\mejo{a_4})^{T}
\nonumber \\
+\mejo{a_1}\xe{1}\mejt{a_2}\ue{2}(\mejt{a_3})^{T}\xe{1}(\mejo{a_4})^{T},~~~~~~~~~~~~~~~
\label{Kredd}
\end{eqnarray}
for the $\alps$ in terms of the normalization%
\index{normalization matrices}
and weight matrices.%
\index{weight matrices}
Conversely, if the coefficients $\alps$ are known then (\ref{Kreaa})
gives $\wejz{a_1}$, and (\ref{Krebb}),(\ref{Krecc}), and (\ref{Kredd})
can be solved successively for $\ue{1},\wejo{a_2},$ and $\ue{2}$.
These enter the equations linearly. Prior to solving each one of
these equations it is necessary to determine the remaining matrices,
$\mejo{a_1},\qejo{a_1},\xe{1},\ve{1}$ or $\mejt{a_2}$ that also
enter the equation in question: these are obtained from their definitions
(\ref{Krooty}), (\ref{Kdefq}), (\ref{Kdefcx}), and (\ref{Kdefv}), 
which give them in terms
of the matrices $\wejz{a_1},\ue{1}$, or $\wejo{a_2},$ found from
solving the previous equations.  
\par
In general the linear equation for the
remaining unknown matrix $\uej$ or $\wej{a}$, encountered at 
respectively the $2j$th stage or $(2j+1)$th stage, will be
sandwiched between products of the matrices $\mei$ and $\xe{i}$.
These linear equations have a solution if we assume, as before,
that the positive semidefinite matrices  $\mei$ and $\xe{i}$
are nonsingular for all $i \le j$. 
\par
It can be checked through matrix
multiplication that the set of matrices  $\gam$ and $\lama{}$
defined via (\ref{Kbigm}) are projection operators satisfying
(\ref{Kproj}), (\ref{Keee}), and (\ref{Klll}) for any choice of normalization
and weight matrices satisfying (\ref{Knaw}). Consequently any further
restrictions on the set of possible normalization and weight
matrices must come from additional information about the
operators  $\gam$ and $\lama{}$, such as the identity (\ref{Kacrp})
which holds for two-dimensional composites. 

Note that we have only shown that the weights and normalization matrices can be recovered
from the coefficients $\alps$. A separate question, which we do not address, is whether these
coefficients can be recovered from the series expansion (\ref{Kself}) in powers of the 
elements of the matrices ${\bfmm{\eps_{a}}}$, $a=1,2,\ldots,p$. Since the matrix ${\bfmm{\eps_{a}}}\lmo$ does not generally commute
with ${\bfmm{\eps_{b}}}\lmo$, when $b\ne a$ it seems likely that one should be able to recover the coefficients $\alps$
if $p$ was sufficiently large. But without a proof the most we can say is what we said in the
introduction: that the series probably contains sufficient information 
to determine the weight and normalization matrices.

\section{Simplification for two-dimensional, isotropic composites}
\setcounter{equation}{0}
It can be checked through matrix
multiplication that the set of matrices  $\gam$ and $\lama{}$
defined via (\ref{Kbigm}) are projection operators satisfying
(\ref{Kproj}), (\ref{Keee}), and (\ref{Klll}) for any choice of normalization
and weight matrices satisfying (\ref{Knaw}). Consequently any further
restrictions on the set of possible normalization and weight
matrices must come from additional information about the
operators  $\gam$ and $\lama{}$, such as the identity (\ref{Kacrp})
which holds for two-dimensional composites.

In particular, if the composite is two-dimensional, statistically isotropic
and has isotropic components then (\ref{Kacrp}) implies that each
normalization matrix is simply the identity matrix. Indeed, 
the isotropy of the composite implies $\efl=\onem\lb^{*}$ for
all choices of moduli $\lba$ and
consequently all the coefficients $\alps$ are also proportional to
$\onem$. It follows that the weights and normalization matrices%
\index{weight matrices}%
\index{normalization matrices}
are also proportional to $\onem$ in their space indices:
\begin{eqnarray}
\wj_{c,a_1...a_s \alpha,b_1...b_s \beta}
=w^{j}_{c,a_1...a_s,b_1...b_s}\delta_{\alpha\beta},
\nonumber \\
\nj_{a_1...a_s \alpha,b_1...b_s \beta}
=n^{j}_{a_1...a_s,b_1...b_s}\delta_{\alpha\beta},
\label{Kisoc}
\end{eqnarray}
and hence commute with $\rpj{}$. The isotropy of the components implies
$\rprperp$ also commutes with the operators $\lama{}$. We next need to
establish that
\beq
\rprperp\xej_{a_1...a_s \alpha}
=(-1)^{s} \sum_{\beta=1}^{2}R^{\perp}_{\alpha \beta}\xej_{a_1...a_s \beta},
\eeq{Karopa}
\beq
\rprperp\yyej{a_1...a_s \alpha}
=\sfac\sum_{\beta=1}^{2}R^{\perp}_{\alpha \beta}\yyej{a_1...a_s \beta}.
\eeq{Karopb}
To see this first observe that (\ref{Kacr}) implies that 
$\cfj{j}$ and $\cgj{j}$ are each closed under the action of $\rprperp$,
and as a consequence so are the spaces $\Xc{j}$ and $\Yc{j}$. Now
we proceed by supposing there exists an $j$ such 
that (\ref{Karopa})  holds true for all $s \le j-1$ and for all permutations
of indices: this is clearly true when $j=0$.
Now for any strings $\rho$ and $\phi$ of length $j-1 \ge 0$, 
(\ref{Kdefyf}) and (\ref{Kdnexx}) imply
\beq \yyej{b\rho} -\sum_{a\ne q} \sum_{\omega} C^{j}_{b\rho,a\omega}(\lama{}\xej_{\omega}) \in \cfj{j-1},
\eeq{Kincaa}
\beq
\xej_{\phi}-\sum_{\tau}D^{j}_{\phi,\tau}(\gam\yyej{\tau}) \in \cgj{j},
\eeq{Kincba}
where $\omega$ has length $j-1$. By our supposition we can
use to (\ref{Karopa}) to compute the action of $\rprperp$ on $\xej_{\omega}$. 
Applying $\rprperp$ to (\ref{Kincaa}) and using (\ref{Kisoc}) and
(\ref{Kacr})  brings one to the conclusion that 
\beq \rprperp\yyej{a_1...a_{j} \alpha}
+(-1)^{j}\sum_{\beta=1}^{2}R^{\perp}_{\alpha \beta}\yyej{a_1...a_{j} \beta}
\in \cfj{j-1},
\eeq{Kinca}
for all combinations of indices. But $\Yc{j}$ is closed under
the action of $\rprperp$ and since $\Yc{j}$ is orthogonal to
$\cfj{j-1}$ we infer that the field in (\ref{Kinca}) is zero, i.e.
that (\ref{Karopb}) holds for $s=m$. Applying $\rprperp$ to
(\ref{Kincba}) and using a similar argument establishes that (\ref{Karopa})
holds when $s=m+1$. By induction this completes
the proof of (\ref{Karopa}) and (\ref{Karopb}). In turn these imply 
via (\ref{Kacrp}) that
\begin{eqnarray}
U^{j}_{\tau,\phi}=(\yyej{\tau},\gam\yyej{\phi})
=(\yyej{\tau},(\rprperp)^{T}\gam\rprperp\yyej{\phi})
=(\yyej{\tau},(I-\gam)\yyej{\phi})
=\delta_{\tau\phi}-U^{j}_{\tau,\phi}.
\nonumber \\
~
\label{Kuhalf}
\end{eqnarray}
From the definition (\ref{Krenu}) it follows that
\beq \norj=\onej, \eeq{Ktwdim}
and consequently the operator $\gam$ is represented by the matrix
\index{representation!g@$\gam$}
\begin{eqnarray}
\gam= \frac{1}{2}\left[ \begin{array}{cccc} 0 & 0 & 0 &\\ 0 & \begin{array}{rr}
\om{1} & \om{1} \\ \om{1} & \om{1} \end{array} & 0 & \\ 0 & 0 &
\begin{array}{rr} \om{2} & \om{2} \\ \om{2} & \om{2} \end{array}
& \\ & & & \ddots \end{array} \right].
\label{Kbiggt}
\end{eqnarray}
Note also from (\ref{Karopa}) and (\ref{Karopb}) that $\rprperp$ 
has the representation,%
\index{representation!r@$\rprperp$}
\begin{eqnarray}
\rprperp=\left[ \begin{array}{ccc} \begin{array}{rr} \rpj{0}~~ & ~\\\
~ & -\rpj{1} \end{array} & 0 & \\ 0 & \begin{array}{rr}
\rpj{1}~~ & ~ \\~& -\rpj{2} \end{array} & \\  &  & \ddots \end{array}
\right],
\label{Kbigrp}
\end{eqnarray}
where $\rpj{j}$ is the rotation matrix with elements
\beq R^{j\bot}_{ a_1...a_s \alpha,b_1...b_s \beta}
=R^{\bot}_{\alpha \beta}\prod_{i=1}^{s}\delta_{a_i b_i}. \eeq{Kderpj}
\par
When $p=2$ and the composite is two-dimensional but possibly
anisotropic, the set of all possible sequences of weight and normalization
matrices has been completely characterized,%
\index{weight matrices!characterization}%
\index{normalization matrices!characterization}
and furthermore microgeometries
have been identified which correspond
to every such sequence. This was accomplished by
\citeAPY{Milton:1986:APLG} for composites of two isotropic phases and 
by \citeAPY{Clark:1994:MEC}  for a polycrystal%
\index{polycrystal}
built from a single
anisotropic crystal. In both cases the microgeometries that can
simulate any sequence were found to be sequentially layered laminates.%
\index{laminate!sequentially layered}
These two-dimensional microstructures can mimic the entire behavior of $\efl$
as a function of the component moduli while keeping the microstructure fixed.
\section{Bounds and methods for bounding the effective tensor}
\setcounter{equation}{0}

Bounds on the effective tensor $\efl$ follow directly from
the variational principles,
\beq \est\cdot\efl\est=\min_{\seb(\rst)}\int_{\Omega}d\rst~(\est+
\seb(\rst))\cdot\lr(\est+\seb(\rst)), \eeq{Kprim}
\beq \jjst\cdot(\efl)^{-1}\jjst=\min_{\sjb(\rst)}\int_{\Omega}d\rst~(\jjst+
\sjb(\rst))\cdot(\lr)^{-1}(\jjst+\sjb(\rst)), \eeq{Kdualp}
where $\est$ and $\jjst$ are uniform fields, and the minimization extends
over statistically homogeneous or periodic fields $\seb(\rst)$ 
and $\sjb(\rst)$ satisfying
\beq \grad\times\seb(\rst)=0 ,~~~~~~~
\int_{\Omega}d\rst~\seb(\rst)=0, \eeq{Kcona}
\beq \diver\sjb(\rst)=0 ,~~~~~~~\int_{\Omega}d\rst~\sjb(\rst)=0. \eeq{Kconb}
Substitution of the trial fields $\seb(\rst)=0$ and $\sjb(\rst)=0$ gives
the arithmetic and harmonic mean bounds,%
\index{bounds!arithmetic and harmonic mean}
\beq [\sum_{a=1}^{p}\als{a}(\lba)^{-1}]^{-1}\le\efls{*j-1}
\le \sum_{a=1}^{p}\als{a}\lba. \eeq{Kelemb}
\par
Better bounds result from a more judicious choice of trial fields. For
example, to derive improved upper bounds one can follow the approach of
Beran (\citeyearNP{Beran:1965:UVA}, \citeyearNP{Beran:1966:UCV}) and choose a trial field of the form
\beq \seb(\rst)=\sum_{s=1}^{j}\sum_{a_1,..,a_s=1}^{p}\sum_{\alp=1}^{3}
\bfmm{c}_{a_1 a_2 ...a_s \alp}\eej_{a_1 a_2 ...a_s \alp}(\rst), \eeq{Ktref}
where the fields $\eej_{a_1 a_2 ...a_s \alp}(\rst)$ are given 
by (\ref{Kdevec}), and then minimize (\ref{Kprim}) to find the best
choice of the coefficients $\bfmm{c}_{a_1 a_2 ...a_s \alp}$,
which are vectors in the field indices. The bound
generated by this procedure when expanded in a power series agrees
with the terms in the series (\ref{Kself}) for all $s$ up to and
including $s=2j+1$, and for this reason is called the Wiener-Beran%
\index{bounds!Wiener-Beran type}
type upper bound of order $2j+1$: a bound is said to be of 
order $m$ if the series expansion of the bound and the
series expansion of $\efl$ agree for all s up to and
including $s=m$. An analogous choice of trial field $\sjb(\rst)$ generates
the Wiener-Beran type lower bound of order $2j+1$ through the variational
principle (\ref{Kdualp}). Bounds of even order are generated by 
substituting an appropriate choice of trial polarization field
into the Hashin-Shtrikman variational principles (\citeAY{Hashin:1962:VAT}).%
\index{variational principles!Hashin-Shtrikman} 
yielding Hashin-Shtrikman type bounds.%
\index{bounds!Hashin-Shtrikman type}

\par
These bounds on $\efl$ are naturally expressed in terms
of the normalization and weight matrices. For this purpose it is useful
to expand $\efl$ as a continued fraction%
\index{continued fraction expansion}
rather than as a power series.
A direct extension of the analysis of Milton (\citeyearNP{Milton:1987:MCEa},\citeyearNP{Milton:1987:MCEb}) 
gives a 
continued fraction expansion for the effective tensor
\beq \efl\equiv\efls{*0}, \eeq{Kstrt}
generated by setting
\beq \lbo=\lb_{q}, \eeq{Ksetlb}
and eliminating the tensors $\efls{*j}$ for $j\ge 1$ from the
recursion relations
\begin{eqnarray}
\efls{*j-1}=\sum_{a=1}^{p}\wejl{a}\lba ~~~~~~~~~~~~~~~~~~~~~~~~~~~~~~~~~~~
~~~~~~~~~~~~~~~~~~~~~~~~~~~~
\nonumber \\
-~\sum_{a,b\ne q}^{}\epa\mej{a}\{\onej\lbo +\sum_{c \ne q}^{}\qej{c}\epc +
(\norj)^{1/2}\efls{*j}(\norj)^{1/2} \}^{-1}(\mej{b})^{T}\epsb,
\nonumber \\
~
\label{Krecur}
\end{eqnarray}
where, in accordance with our previous definitions,
\begin{eqnarray}
\qej{c}=\mej{c}(\wejl{c})^{-1}(\mej{c})^{T},~~~\mej{}=(\yej)^{1/2},~~~
\yj_{a\omega,b\rho}=\delta_{ab}\wjl_{a,\omega,\rho}
-\sum_{\zeta}\wjl_{a,\omega,\zeta}\wjl_{b,\zeta,\rho},
\nonumber \\
~
\label{Krecal}
\end{eqnarray}
and $\mej{a}$, with transpose $(\mej{a})^{T}$, is the rectangular 
submatrix of the square matrix $\mej{}$ 
with elements $\mj_{a\tau,\lambda}$ labeled by
the strings $\tau$ and $\lambda$.
Note that $\efls{*j}$ has elements
$\lbsj_{\tau k \eta m}$ labeled by field indices $k,m \in \{1,2..p\}$
and string indices $\tau=a_{1}a_{2}..a_{j}\alpha$,$~\mu=b_{1}b_{2}..b_{j}
\beta$ with $a_{i}$ and $b_{i}~\in \{1,2,..q-1,q+1,..p\}$, and
$\alpha$ and $\beta \in \{1,2,3\}$. Also note 
that $\wej{a},\mej{a}$
and $\norj$ act on the string indices, not on the field indices.
\par
	There are other equivalent ways of expressing $\efls{*j-1}$ in
terms of $\efls{*j}$ (\citeAY{Milton:1987:MCEa}). For 
example (\ref{Krecur}) can be replaced by its dual form
\begin{eqnarray}
(\efls{*j-1})^{-1}=\sum_{a=1}^{p}\wejl{a}(\lba)^{-1} ~~~~~~~~~~~~~~~~~~~~~~~~~~~~~~~~~~~~~~~~~~~~~~~~~~~~~~~
\nonumber \\
-~\sum_{a,b\ne q}^{}\npa\mej{a}\{\onej\lmo +\sum_{c \ne q}^{}\qej{c}\npc +
(\norj)^{-1/2}(\efls{*j})^{-1}(\norj)^{-1/2} \}^{-1}(\mej{b})^{T}\npb,
\nonumber \\
~
\label{Krecin}
\end{eqnarray}
where 
\beq \npa \equiv (\lba)^{-1}-\lmo. \eeq{Kdenpc}
Eliminating the matrices $\efls{*j}$ from this recursion relation
generates an alternative continued fraction expansion%
\index{continued fraction expansion}
of $\efl$.
\par
The tensors $\efls{*j}, j=0,1,2,...$ have 
an interpretation in the context of the solution $\jrst$ for
any given field $\est \in \Xc{j}$ (the space $\Xc{j}$ now plays
the role that was played by the uniform fields) to the
equations
\beq \gamj\jb=0,~~\jrst=\lr(\est(\rst)+\seb(\rst)),
~~\gamj\seb=\seb,\eeq{Kgenr}
where $\gamj$ is the nonlocal operator,
\beq \gamj=\gam-\upsj, \eeq{Kmogam}
and $\upsj$ (which commutes with $\gam$) is the projection onto the
space
\beq {\cal{E}}^{j} \equiv 
\{ \vec{u}(\rst) \in \Xc{j}\oplus\Yc{j} | \gam\vec{u}=\vec{u} \}
\eeq{Kdefxsp}
of order $j$ fields which are curl-free and have zero average value. In
the representation (\ref{Kbigm}) $\gamj$ is obtained from $\gam$ by
setting the blocks $\ue{j},\ve{j}$ and $\xe{j}$ to zero. Note that 
$\gamj$ is a projection and acts upon any field to produce 
a curl-free field with zero average value. So in particular
$\seb(\rst)$ (but not $\est(\rst)$) is the gradient of a potential. 
\par
A simple application of the Lax-Milgram lemma%
\index{Lax-Milgram lemma}
(see, for example, Section 5.8
of \citeAY{Gilbarg:2002:SOH}) shows that
these equations always have a unique solution for $\jrst$, for 
any choice of field $\est \in \Xc{j}$, provided that the
set of tensors $\lama{}$ are positive definite and bounded. Let us
define $\gamo^{j}$ as the projection onto the 
subspace $\Xc{j}$ and $\jjst$ as the component 
\beq \jjst=\gamo^{j}\bfmm{J}, \eeq{Kprojj}
of the field $\jrst$ which lies in the subspace $\Xc{j}$. Since the
relation between $\jjst$ and $\est$ is linear we can write 
\beq \jjst=\efls{*j}\est. \eeq{Kdelsj}
This linear relation serves to define  $\efls{*j}$: it is a linear map
from the space $\Xc{j}$ to itself. When $j=0$ these equations 
reduce to the previous set (\ref{Kconst}), (\ref{Ksmalle}), 
and (\ref{Krela}) and so we can make the identification (\ref{Kstrt})
between $\efl$ and $\efls{*0}$.
\par
From the matrix representation (\ref{Kbigm}) of the operators
$\lama{}$ and $\gam$ it is clear that $\lama{}$ and $\gamj$ do not
couple $\est$ with fields in the space $\cgj{j}$. Thus the
fields in $\cgj{j}$ play no role in the solutions of the equations 
(\ref{Kgenr}). Consequently we can now eliminate from our basis those fields
$\xej_{\tau},\yyej{\tau} \in \cgj{j}$. In the remaining reduced basis
the operators $\lama{}$ and $\gamj$ have the representation%
\index{representation!projection operators}%
\index{projection operators!representation}
\begin{eqnarray}
\lama{} = \left[ \begin{array}{ccc} \begin{array}{rr} \wea{j}~~ & \mea{j+1}\\
(\mea{j+1})^{T} & \qea{j+1} \end{array} & 0 & \\ 0 & \begin{array}{rr}
\wea{j+1}~~ & \mea{j+2} \\ (\mea{j+2})^{T}& \qea{j+2} \end{array} & \\  &  & \ddots \end{array}
\right],
\nonumber \\
\gamj= \left[ \begin{array}{cccc} 0 & 0 & 0 &\\ 0 & \begin{array}{rr}
\ue{j+1} & \xe{j+1} \\ \xe{j+1} & \ve{j+1} \end{array} & 0 & \\ 0 & 0 &
\begin{array}{rr} \ue{j+2} & \xe{j+2} \\ \xe{j+2} & \ve{j+2} \end{array}
& \\ & & & \ddots \end{array} \right].~~~~~
\nonumber \\
\label{Kmbigm}
\end{eqnarray}
The similarity with (\ref{Kbigm}) makes it evident that whatever role
the sequence $\wea{0}$,$\nm{1}$,$\wea{1}$,$\nm{2}$,$\wea{2},...$ of
weight and normalization matrices plays in
determining $\efl$ is played in an identical way by the sequence
$\wea{j}$,$\nm{j+1}$,$\wea{j+1}$,$\nm{j+2}$,$\wea{j+2},...$ in
determining $\efls{*j}$. This self-similarity is also evident
from the continued fraction expansions for $\efl$ and $\efls{*j}$
implied by (\ref{Krecur}).

\par
If the entire set of normalization and weight matrices is known
then these continued fractions expansions allow the effective tensor
$\efl$ to be computed to an arbitrarily high degree of accuracy. For 
example we could truncate%
\index{continued fraction expansion!truncation}
the continued 
fraction at some stage $m$ by setting
\beq \efls{*m}=\onejm\lb_{q}, \eeq{Kaprox}
which is a natural choice, corresponding to replacing the set of weights
$\wea{m}$ by the weights
\beq \weq = \onejm,~~~~\wea{m}=0,~~\forall a\ne q, \eeq{Kredef}
consistent with the constraints (\ref{Knaw}). Then the tensor $\efls{*0}$
obtained from the recursion relations (\ref{Krecur}) 
is an $m$-th order rational approximate to $\efl$, and it can be proved
that this approximate converges to $\efl$ as $m$ tends to infinity, for
any positive definite bounded set of moduli  $\lba, a=1,..,p$ (\citeAY{Milton:1987:MCEb}).
The approximates also converge when the moduli are complex,  
provided the tensors $\lba$ are symmetric and bounded and such that
there exists a phase angle $\theta$ for which
\beq \Real(e^{i\theta}\lba) > 0,~~~~~\forall a, \eeq{Kcomp}
where $\Real(A)$ denotes the real part of the quantity $A$. Such complex
moduli have a physical interpretation. When the fields $\jb$ and
$\ebvec$ oscillate sinusoidally in time $t$ 
with frequency $\omega$ then
they can be expressed as the real part of complex fields $\jb_{c}(\rst)$
and $\ebvec_{c}(\rst)$,
\beq \jb(\rst,\omega)=\Real(e^{iwt}\jb_{c}(\rst)),
~~~~\ebvec(\rst,\omega)=\Real(e^{iwt}\ebvec_{c}(\rst)). \eeq{Koscil}
Provided the wavelength of this oscillation is sufficiently large 
compared with the microstructure these complex fields satisfy the
quasistatic equations,
\beq \diver\jb_{c}(\rst)=0,~~~\grad\times\ebvec_{c}(\rst)=0,~~~
\jb_{c}(\rst)=\lr\ebvec_{c}(\rst), \eeq{Kqstat}
with a complex tensor $\lr$  given by
\beq \lr=\sum_{a=1}^{p}\lama{}\lba, \eeq{Kdecoml}
where the moduli $\lba$ are complex and frequency dependent. 
The thermodynamic requirement that dissipation of power into entropy%
\index{entropy}
be positive
ensures that (\ref{Kcomp})  holds when $\theta=0$.
Each rational approximate satisfies the properties of covariance%
\index{covariance}
and disjunction,%
\index{disjunction}
discussed in the introduction, and has the additional
required analytic property that 
\beq \Real(e^{i\theta}\efl) > 0, \eeq{Kana}
for any set of tensors $\lba$ satisfying (\ref{Kcomp}).
\par
	Bounds on $\efl$ follow from elementary bounds%
\index{bounds!elementary}
on $\efls{*j}$.
In particular, the inequalities
\beq 0 \le \efls{*j} \le \infty\onej, \eeq{Kbsimp}
or equivalently the inequalities

\beq [\sum_{a=1}^{p}\wejl{a}(\lba)^{-1}]^{-1}\le\efls{*j-1}
\le \sum_{a=1}^{p}\wejl{a}\lba, \eeq{Kwber}
when substituted in the recursion relations (\ref{Krecur}) 
or (\ref{Kdenpc}) produce the
Weiner-Beran type bounds%
\index{bounds!Weiner-Beran type}
on $\efl$ of order $2j-1$, while the inequalities
\beq \lb^{-}\onej\le\efls{*j}\le\lb^{+}\onej, \eeq{Khash}
which hold for all tensors $\lb^{-}$ and $\lb^{+}$ such that
\beq \lb^{-}\le\lba\le\lb^{+},~~~~1\le a \le p ,\eeq{Kdref}
when substituted in (\ref{Krecur}) 
or (\ref{Kdenpc}) produce the Hashin-Shtrikman type bounds%
\index{bounds!Hashin-Shtrikman type}
on $\efl$ of order $2j$. By substitution we mean precisely that an upper
bound on $\efl$ is obtained by setting $\efls{*j}=\infty\onej$ or
$\efls{*j}=\lb^{+}$ and solving the recursion relations
for $\efls{*0}$ and that a lower bound on $\efl$ is obtained by setting
$\efls{*j}=0$ or $\efls{*j}=\lb^{-}$ and solving for $\efls{*0}$.
\section{Bounds using the field-equation recursion method}
\labsect{Kfrm}
\setcounter{equation}{0}
The inequalities (\ref{Kwber}) and (\ref{Khash}) can be easily 
derived without reference to variational principles using the field
recursion method for bounding effective tensors. This approach utilizes the
recursive structure of the equations (\ref{Krecur}) and
the inequalities (\ref{Knaw}) on the normalization and weight matrices.
The first step in the method is 
to conjecture a set of restrictions that might apply to
$\efls{*j}$ irrespective of what values the weights and normalization
matrices take, subject only to the constraints (\ref{Knaw})-or perhaps
additional constraints if these are known. 
This conjecture need not be very restrictive, and could be guided by
the form of the recursion relations (\ref{Krecur}).
For example let us conjecture that $\efls{*j}$ is positive
semidefinite. The next step is to first check that the tensor
$\efls{*m}$ given by (\ref{Kaprox}) satisfies the conjecture, and indeed it
does. Then the remaining task is to 
assume the conjecture is true for some $j$ 
and show this implies $\efls{*j-1}$ also satisfies the conjecture, for any
choice of the weight matrices $\wejl{a}$ and normalization 
matrices $\norj$ satisfying (\ref{Knaw}): it obviously does since from
the recursion relations (\ref{Krecur}) and (\ref{Krecin}) it follows that
(\ref{Kbsimp}) implies (\ref{Kwber}) which in turn implies $\efls{*j-1}$
is positive semidefinite.
By induction any rational approximate for $\efls{*j}$ generated
by choosing $m>j$ and making the substitution (\ref{Kaprox}) satisfies 
the conjecture,
and since these approximates converge to $\efls{*j}$ 
as $m$ tends to infinity, we conclude that $\efls{*j}$ itself must be
positive semidefinite.
The conjecture is proved and it clearly implies both (\ref{Kwber}) and
(\ref{Khash}).  The recursion method has the advantage that it also
works when the moduli $\lba$ are complex (Milton, \citeyearNP{Milton:1987:MCEa}; \citeyearNP{Milton:1987:MCEb})
\par
In the special case of a composite with $p=2$ the 
strings of indices merely consist of a repeated string of either $2's$
or $1's$ (according to whether $q=1$ or $q=2$) terminated by a space
index. Let us drop this redundant information and allow the elements of
the weight and normalization matrices%
\index{weight matrices}%
\index{normalization matrices}
to be addressed only by the 
space indices. Also when $p=2$ the matrices
$\wej{1}$ and $\wej{2}=\onej-\wej{1}$ commute and so we have
\beq \yej=\wej{1}\wej{2} ,~~~~~~\mej{}=(\wej{1}\wej{2})^{1/2},~~~~~~
\qej{1}=\wej{2},~~~~~~ \qej{2}=\wej{1}. \eeq{Ktcomp}
Without loss of generality we take $q=2$, and 
correspondingly $\lbo=\lb_{2}$.
Then the recursion relation (\ref{Krecur}) simplifies to
\begin{eqnarray}
\efls{*j-1}=\wejl{1}\lb_{1}+\wejl{2}\lb_{2} ~~~~~~~~~~~~~~~~~~~~~~~~~~~~~~~~~~~~~~~~~~~~~~~~~~~~~~~~~~~~~~~~~~~
\nonumber \\
-~(\lb_{1}-\lb_{2})\mej{}\{\wejl{1}\lb_{2}+\wejl{2}\lb_{1}+
(\norj)^{1/2}\efls{*j}(\norj)^{1/2} \}^{-1}\mej{}(\lb_{1}-\lb_{2}),
\nonumber \\
~
\label{Krcurt}
\end{eqnarray}
which for $\lb_{1} \ne \lb_{2}$ can be inverted to give $\efls{*j}$ in 
terms of $\efls{*j-1}$:
\begin{eqnarray}
\efls{*j}=(\norj)^{-1/2}\{-\wejl{1}\lb_{2}-\wejl{2}\lb_{1} ~~~~~~~~~~~~~~~~~~~~~~~~~~~~~~~~~~~~~~~~~~~~~~~~~~~~~
\nonumber \\
 +~(\lb_{1}-\lb_{2})\mej{}(\wejl{1}\lb_{1}+\wejl{2}\lb_{2}
-\efls{*j-1})^{-1}\mej{}(\lb_{1}-\lb_{2})\}(\norj)^{-1/2}.
\nonumber \\
\label{Kirecur}
\end{eqnarray}
\par
Supposing that the components are 
isotropic phases occupying volume fractions $f_1$ and $f_2$, (\ref{Ktra})
implies
\beq W^{0}_{1,\alpha,\beta}=f_{1}\delta_{\alpha\beta},~~~~~~~~
 W^{0}_{2,\alpha,\beta}=f_{2} \delta_{\alpha\beta}, \eeq{Kwetin}
and consequently when $j=1$ (\ref{Kirecur}) takes the form
\beq \efls{*1}=({{\stackrel{\leftrightarrow}{N}}{}^{1}})^{-1/2}\yest{*}({{\stackrel{\leftrightarrow}{N}}{}^{1}})^{-1/2}, \eeq{Kystar}
where $\yest{*}$, not to be confused with the matrix $\yej $, is given by
\begin{eqnarray}
\yest{*}=-f_{1}\onem\lb_{2}-f_{2}\onem\lb_{1} 
 +~f_{1}f_{2}(\lb_{1}-\lb_{2})(f_{1}\onem\lb_{1}
+f_{2}\onem\lb_{2}-\efl)^{-1}(\lb_{1}-\lb_{2}).
\nonumber \\
 ~
\label{Kiirec}
\end{eqnarray}

\section{Bounds using the translation method}
\setcounter{equation}{0}
It turns out that bounds on $\efl$ derived via the translation method%
\index{translation method} 
follow from elementary bounds%
\index{bounds!elementary}
on this tensor $\yest{*}$.
This method was discovered independently by Murat and Tartar 
(\citeyearNP{Tartar:1979:ECH};\citeyearNP{Murat:1985:CVH};\citeyearNP{Tartar:1985:EFC}) 
and by Lurie and Cherkaev (\citeyearNP{Lurie:1982:AEC};\citeyearNP{Lurie:1984:EEC}) 
and applied to generate bounds that characterize for $n=1$
the region in tensor space filled by the range of values $\efl$
takes as the microstructure varies over all configurations 
while keeping the moduli $\lb_{1}$ and $\lb_{2}$
and the volume fraction $f_{1}$ fixed. Subsequently
it was noted that the corresponding region filled by the possible
values of $\yest{*}$ did not depend on the choice of
volume fraction $f_{1}$ (\citeAY{Milton:1986:MPC}). 
\citeAPY{Cherkaev:1992:ECB} extended the characterization to $n=2$,
assuming a two-dimensional geometry. Subsequently \citeAPY{Clark:1995:OBC}
obtained the characterization for arbitrary $n$, using fractional linear transformations%
\index{fractional linear transformation}
which preserve the analytic properties as functions of the component moduli.
\index{analyticity in component moduli}

To explain the translation
method let us focus on bounding $\efl$ from below. Then  
one needs to  find a suitable
translation tensor $T_{\alpha i\beta k}$, where $i,k$ are field indices
and $\alpha,\beta$ are space indices, satisfying
\beq \onem\lba~ \ge~ \tes, ~~~~a=1,2, \eeq{Ktesieq}
and with the additional property that 
\beq \int_{\Omega} d\rst~\grad\psib\cdot\tes\grad\psib \ge 0, \eeq{Ktrans}
for all periodic potentials $\psib$
with elements $\psi^{k}(\rst), k=1,2,..n$.
Any positive semidefinite tensor satisfies this last constraint.
However the converse is not true, and in fact
the interesting applications to bounds come from translations
$\tes$ which are not positive semidefinite. The key idea in
the method is to consider a comparison composite with its
moduli translated from $\lr$ to the moduli
\beq \lrp \equiv \lr-\tes, \eeq{Kcompa}
which are positive semidefinite as a consequence of (\ref{Ktesieq}).
From (\ref{Ktrans})
and from the variational definition (\ref{Kprim}) applied to 
the effective tensor
$\eflp$ of the comparison composite we have, for all
uniform fields $\est$,
\begin{eqnarray}
\est\cdot\eflp\est=\min_{\psib(\rst)}\{\int_{\Omega}d\rst~(\est-
\grad\psib(\rst))\cdot(\lrp)(\est-\grad\psib(\rst))\}~~~~~~~~~~~~~~~~
\nonumber \\
=\min_{\psib(\rst)}\{\int_{\Omega}d\rst~(\est-
\grad\psib(\rst))\cdot(\lr)(\est-\grad\psib(\rst))~~~~~~~~~~~~~~~~
\nonumber \\
-\int_{\Omega}d\rst~(\grad\psib(\rst))\cdot(\tes)(\grad\psib(\rst))
~-\est\cdot\tes\est\}
\nonumber \\
~\le \min_{\psib(\rst)}\{\int_{\Omega}d\rst~(\est-
\grad\psib(\rst))\cdot(\lr)(\est-\grad\psib(\rst))\}
~-\est\cdot\tes\est
\nonumber \\
~= \est\cdot(\efl-\tes)\est, ~~~~~~~~~~~~~~~~~~~~~~~~~~~~~~~~~~~~~~~~~~~~~~~~~~~~~~~
\label{Kinseq}
\end{eqnarray}
which is equivalent to the tensor inequality 
\beq \eflp~\le~\efl-\tes. \eeq{Ktineq}
Substituting this in the harmonic mean bounds%
\index{bounds!harmonic mean}
on $\efl$,
\beq (\eflp)^{-1}~\le~\int_{\Omega}d\rst~(\lrp)^{-1}, \eeq{Kharmo}
yields the translation bounds,
\beq (\efl-\tes)^{-1}~\le~\int_{\Omega}d\rst~(\lr-\tes)^{-1}, \eeq{Ktharm}
which for composites of two isotropic materials reduces to
\beq  (\efl-\tes)^{-1}~\le~ f_{1}(\onem\lb_{1}-\tes)^{-1}
 +f_{2}(\onem\lb_{2}-\tes)^{-1}. \eeq{Kinvav}

\citeAPY{Cherkaev:1992:ECB}
noticed through algebraic manipulation, that these bounds when
expressed in terms of $\yest{*}$ simplify to
\beq \yest{*}+\tes~ \ge~ 0. \eeq{Ksimtr}
In their proof they assumed that $\lb_{1}$ and $\lb_{2}$ commute.
Later this assumption was found unnecessary and moreover
a direct and simple proof of (\ref{Ksimtr}) was found from a variational
expression for $\yest{*}$ (\citeAY{Milton:1991:FER}). An interesting feature of the
translation method is that the sharpest bounds are usually
obtained from translations $\tes$ with couplings
between the fields, even when $\lb_{1}$ and $\lb_{2}$, and hence $\efl$,
have no such couplings. 
\par
When $p>2$ the transformation (\ref{Krecur}) cannot simply be inverted
because the matrices $\mej{a}$ are rectangular and have no unique
inverse. Also it is clear that the tensor $\efls{*j}$ is larger than
the tensor $\efls{*j-1}$ and so contains more information. However
if more than one field was present, i.e. if $n \ge 2$, and 
if $\efl$ was known as a function of the $\lba$, then in principle
one could expand $\efl$ in a power series, possibly extract the coefficients
$\alps$ and subsequently find the weights and normalization matrices.
By this means one could recover both $\efls{*j}$ and 
\beq \yest{*j} \equiv (\norj)^{-1/2}\efls{*j}(\norj)^{-1/2} \eeq{Kdefyjs}
as a function of the $\lba$
through the continued fraction formula for $\efls{*j}$ implied by
($\ref{Krecur}$). Naturally we expect that there 
exists a more direct way of 
recovering  the function $\yest{*j}(\lb_1,\lb_2,..,\lb_p)$ from the
function $\efls{*j-1}(\lb_1,\lb_2,..,\lb_p)$. 
One intriguing question is whether this direct recovery process,
whatever it is,
works when $n=1$. If it does then the sequence of matrices
$\norj$ and $\wej{a}$ could be recovered by expanding 
each function $\yest{*j-1}(\lb_1,\lb_2,..,\lb_p)$ to first order, and
consequently $\efl$ could be calculated even when more than one
field is present. In other words, knowledge of the conductivity
function $\sigs(\sia{1},\sia{2},...\sia{p})$
without couplings would be sufficient to uniquely determine
the effective tensor $\efl$ with couplings present.
\section*{Acknowledgments}
This manuscript was largely complete in 1992, and at that time Graeme W. Milton was an associate professor at the Courant Institute,
and a recipient of a Packard Fellowship, and Mordehai Milgrom was a visitor to the Courant Institute. 
Conversely Graeme W. Milton benefited from a visit to the
Weizmann Institute. The Courant Institute, the Packard Foundation, the University of Utah, and the Weizmann Institute are gratefully thanked for their support.

\bibliographystyle{mod-xchicago}
\bibliography{/home/milton/tcbook,/home/milton/newref}

\end{document}